\documentstyle[12pt]{article}
\newtheorem{thm}{Theorem}

\newtheorem{lem}{Lemma}
\newtheorem{df}{Definition}

 \newcommand{\Integer}{\:\mbox{\sf Z} \hspace{-0.82em} \mbox{\sf Z}\,}
  
 \newcommand{\Real}{\mbox{I \hspace{-0.82em} R}}
 
 \newcommand{\Complex}{
        \mbox{C \hspace{-1.16em} \raisebox{-0.018em}{\sf l}}\;}
 \newcommand{\Z}{\Integer}

 \newcommand{\GL}{gap labelling}
 \newcommand{\IDS}{integrated density of states}
 \newcommand{\tat}{t}
 
 \newcommand{\bd}{\begin{df}}
 \newcommand{\bt}{\begin{thm}}
 \newcommand{\bl}{\begin{lem}}
 \newcommand{\ed}{\end{df}}
 \newcommand{\be}{\begin{equation}}
 \newcommand{\ee}{\end{equation}}
 \newcommand{\bi}{\begin{itemize}}
 \newcommand{\ei}{\end{itemize}}
 \newcommand{\et}{\end{thm}}
 \newcommand{\el}{\end{lem}}
 \newcommand{\bs}{\bigskip}
 \newcommand{\ms}{\medskip}
 \newcommand{\eb}{\hfill$\Box$}
 \newcommand{\tl}{{\cal T}}
 
 \newcommand{\tr}{\hbox{tr}}
 \newcommand{\pkt}{puncture}
 \newcommand{\Tl}{{\cal K}}
 \newcommand{\K}{M_\infty(\Complex)}
 \newcommand{\Sy}{G}
 \newcommand{\Ch}{\chi^{}}
 \newcommand{\A}{{\cal A}}
 \newcommand{\2}{^{pct}}
 \newcommand{\mi}{pattern}
 \newcommand{\mii}{oriented pattern class}
 \newcommand{\miii}{pattern class}
 \newcommand{\ti}{tile}
 \newcommand{\tii}{oriented prototile}
 \newcommand{\tiii}{prototile}

 \newcommand{\B}{{\cal B}}
 \newcommand{\bii}{\B2}
 
 \newcommand{\bo}{\B1}
 
 \newcommand{\jo}{\Xi}
 \newcommand{\jk}{Y}
 \newcommand{\joh}{\tilde{\jo}}
 
 \newcommand{\k}{\kappa}
 \newcommand{\LI}{{\cal L}_{\tl}}
 
 \newcommand{\Om}{\Omega}
 
 \newcommand{\Ga}{{\cal G}}
 
 \newcommand{\pf}{{\cal P}_{\Sigma}}
 
 \newcommand{\AF}{{\cal A}_{\Sigma}}
 \newcommand{\p}[2]{{\cal P}^{(#1)}_{\Sigma,#2}}
 
 \newcommand{\mal}[2]{\!\times_{#1}\!\Integer^{#2}}
 \newcommand{\AT}[1]{\A_{\tl_{#1}}}
  \newcommand{\kl}[2]{\xi^{(#1)}_{<,#2}}
 \newcommand{\gr}[2]{\xi^{(#1)}_{>,#2}}
 \newcommand{\pfw}{Perron-Frobenius-eigen\-value}
 \newcommand{\pfv}{Perron-Frobenius-ei\-gen\-vec\-tor}
 \newcommand{\Gr}{{\cal R}}
 \newcommand{\Gru}{\Gamma}
 \newcommand{\So}{Schr\"odinger operator}
 
 \newcommand{\cm}{connectivity matrix}
 \newcommand{\sst}{substitution}
 \newcommand{\dfl}{deflation}
 \newcommand{\ifl}{inflation}
 \newcommand{\fl}{substitute}
 \newcommand{\nc}{non commutative}
 \newcommand{\ncg}{non commutative geometry}
 \newcommand{\saum}{border}
 \newcommand{\CA}{$C^*$-algebra}
 
 \newcommand{\kg}{$K$-group}
 \newcommand{\kog}{$K_0$-group}
\newcommand{\ap}{$\rho$-chain}
 \newcommand{\Mi}[2]{M_{#1}(#2)}
 \newcommand{\pk}[2]{x_{#1}(#2)}

 \title{
 \bf Non Commutative Geometry of Tilings\\ and Gap Labelling
   \vspace{1.5em}}
 \author{Johannes Kellendonk\thanks{email: johannes@mth.kcl.ac.uk}
         \vspace{1em}}
 \date{Department of Mathematics, King's College London,\\ Strand,
 London WC2R 2LS}

\begin{document}
 \maketitle

 \begin{abstract}
\noindent
To a given tiling a \nc\ space and the corresponding
\CA\ are constructed.
This includes the definition of a topology on the groupoid induced by
translations of the tiling.
The algebra is also the algebra of observables for discrete
models of one or many particle systems on the tiling or its periodic
identification. Its scaled ordered
\kog\ furnishes the \GL\ of \So s. The group is computed
for one dimensional tilings and Cartesian products thereof. Its image under
a state is investigated for tilings which are
invariant under a \sst.
Part of this image is given by an invariant measure
on the hull of the tiling which is determined.
The results from the Cartesian products of one dimensional tilings point out
that the gap labelling by means of the values of the \IDS\ is already
fully determined by this measure.

 \end{abstract}

 \begin{flushright}
 \parbox{12em}
  { \begin{center}
      KCL-TH-94-1
 \end{center} }
 \end{flushright}

\newpage

\bibliographystyle{unsrt}

\tableofcontents

\section*{Introduction}

\addcontentsline{toc}{section}{\bf Introduction}

The spatial structure of crystal solids is usually described by point lattices,
with each point representing an elementary cell.
This cell encodes the local arrangement of the atoms.
To get insight into the physical properties of the crystal one may study
discrete models in which a finite
dimensional Hilbert space is assigned to each elementary cell. In particular
the typical features of particle spectra
arrising from translational symmetry already show up in these models.
More generally the spatial structure of a solid may be described by a tiling
(its three dimensional analog), which
 is a covering of the Euclidean space with
tiles which do not overlap.
In this more general setting, too,
it is expected that characteristic features of the solid which depend on the
spatial structure show up in the analysis of discrete models  over the tiling
provided there is some sort of order in the tiling.

The most famous tilings lacking translational
symmetry are the non periodic self-similar ones where in many cases
the order is manifested in a \sst\ procedure, which is a kind of
self-similarity transformation.
Amongst these the Penrose tilings \cite{Pen} are very well known
not only because they have nice mathematical properties but also since
a three dimensional
generalization of them \cite{Gar,KrNe,LeSt1}
serves as an idealized model for certain quasicrystals, which have first
been observed in 1984 \cite{She}.
Quasicrystals
have a long range orientational order with a point symmetry which is not
compatible with a periodic structure. Up to now quasicrystals with
8-, 10-, and 12-fold symmetry have been observed and correspondingly
the interest in tilings with that kind of symmetry is particularly high.
\ms

The present work concerns the \ncg\ of a tiling and as a direct application
the \GL\ of discrete \So s of quantum mechanical systems on the
tiling.
Gap labelling is meant here as a characterization of the spectrum of the
\So\ which does not take the specific properties (such as the strengths
of the bindings) into account but rather the order in the solid.
It is based on the expectation
that
the spatial structure of the tiling
influences the spectrum if it is implemented in the
potential in a very general way,
namely the value of the potential on a tile shall only
depend on the local pattern of tiles around it.
If, for instance, the tiling is self-similar and non periodic one typically
expects a singular continuous spectrum which is a Cantor set of measure zero
and which is self-similar, too, c.f.\ \cite{BG} and references therein
as well as related articles from \cite{OsSt}.
Rather generally speaking \GL\ is an the assignment of labels, as elements of
a countable set, to the gaps of a \So. A useful candidate for this set is
given by the values of the \IDS, since it is an ordered set and allows to
assign a relative location in the spectrum to different gaps.
Also here the spatial structure shows up. Whereas periodic tilings lead to
 a finite set of gaps, the non periodic self-similar ones are expected to
yield integrated densities of states with the values on gaps being dense in
$[0,1]$ and forming a self-similar set.
Being interested in the generic features of the spectrum
which depend on the spatial structure only
we shall not choose a specific operator and
compute its spectrum (which might in fact be to difficult to carry out)
but rather use the abstract form of the \GL\ and construct
an Abelian group from the \ncg\ of the tiling.
The gap labelling is then a map from the gaps of a \So\ into that group.
This group is the scaled
ordered \kog\
(or its image under a homomorphism) of the \CA\ which is constructed from the
tiling and which may be understood as the algebra of observables. It
should at best be in bijective
correspondence to the gaps of a \So.\ms

The above-mentioned self-similar tilings may also be of interest in other
areas of physics where self-similarity is of importance, e.g.\ in conformal
field theory. Their structure may be analysed by methods of \ncg. In fact,
the \nc\
space of a Penrose tiling is the first example in A. Connes book \cite{Con}
for the new concept of space. In this example Connes uses a map found by
Robinson \cite{GrSh} from the set of all Penrose tilings to the set of
$0,1$-sequences satisfying the constraint
$\forall n: a_n+a_{n+1}\leq 1$ to construct a \nc\ space and the corresponding
\CA\ of a Penrose tiling. This has been the guiding line of our investigations,
although the algebra obtained in \cite{Con}
does not contain all translation operators and consequently also not the
discrete Laplacian.
A more careful look at the equivalence relation induced by translation
leads to algebras which contain the discrete Laplacian.
However they are not finitely approximated and their ordered \kog s are not
in all interesting cases (including the Penrose tilings) known to us.
In order to obtain at least part of the \GL\ we have to rely on the
$AF$-algebra
which is determined by the self-similarity structure.
It allows us in particular to determine a measure for
 the frequency of
patterns in the tiling. For Cartesian products of
 one dimensional tilings this measure
already determines the \GL\ by means of the values of the \IDS, and
we expect this to hold for more general tilings, too.\bs

The first section contains the definition of a tiling and its role as an
underlying space for a discrete model of a solid.
Furthermore the \GL\ as
developed by Bellissard \cite{Be1,Be2} is
briefly explained.

The \ncg\ of a tiling is discussed in section two.
The \nc\ space of a tiling is the quotient of its hull by the equivalence
relation induced by translations of a tiling as a whole.
It is non Hausdorff in the non periodic case.
The algebra $\A_\tl$ of the tiling $\tl$ is the
groupoid-\CA\ of the groupoid defined by the equivalence relation.
For its definition a topology on the groupoid is introduced
which is of subtle importance and which is well known
if the tiling can be identified with a mapping from a discrete group into the
set of (congruence classes of) tiles.
The invariants of $\A_\tl$,
e.g.\ the \kg s, may be used to characterize the tiling
and to distinguish it from others.

As a first application some general considerations about one dimensional
tilings and Cartesian products thereof are made. We are mainly interested in
non periodic tilings and just briefly touch on periodic ones. The results
concerning the algebra and \kg s of Cartesian products may be expressed
entirely in terms of the one dimensional components.

To obtain more concrete results and treat also tilings having
$8$-, $10$-, and $12$-fold orientational symmetry we restrict in the fourth
section to tilings which are invariant under a \sst.
This is used to construct the analog of a Robinson map $\jo$ which
provides us with an $AF$-algebra $\AF$ also naturally assigned to the tiling.
It allows us to determine the invariant measure on the hull of a tiling which
measures the relative frequency of patterns. Two different kind of
generalizations of the Fibonacci chain and the Penrose tilings follow as
examples.

In the last section we compare the two \CA s obtained. This not only sheds
some light onto the role of symmetry axes which appear as an obstruction to
approximating the Laplacian by elements of $\AF$ in the von Neumann topology
but also allows us to prove that Shubin's formula holds for the tilings under
investigation. We include a paragraph on \sst\ sequences (one dimensional
\sst\ tilings), for which the ordered \kog s of $\A_\tl$ and $\AF$ coincide.
This is shown by establishing the Vershik transform \cite{Ver,HPS} which turns
out
to be particularly simple if constructed using the path space determined by
the \sst.\bs

\section{Gap labelling of \So s on tilings}

A $d$ dimensional tiling\footnote{
Abusing the language we shall call higher and lower dimensional analogs of a
 tiling or a tile by the same name.}
$\tl$ of $\Real^d$ is a complete covering of $\Real^d$ by \ti s which
do not overlap. A \ti\  shall here be polyhedron in which a point is
distinguished which we call the \pkt\ of the tile.
Occasionally we allow a tile to be marked, e.g.\ by arrows, to break
its symmetries.
The congruence class of a \ti\ under Euclidean transformations is usually
called a \tiii\ and the term \tii\ will be used for an equivalence class
under translations only. Several \ti s of a tiling make up a \mi.
 Correspondingly there are \miii es, i.e.\ congruence classes under all
Euclidean transformations, and \mii es which are equivalence classes
under translations only.
The set of \pkt s of $\tl$ is a subset of $\Real^d$ and will be denoted by
 $\tl\2$.
The set of \pkt s of an \mii\
$M$ is denoted by $M\2$, and it may be identified with a subset of $\Real^d$
once a representative for $M$ has been chosen.
In fact
$(M,x)\subset (T,y)$ shall denote the representative of the \mii\ $M$
which occurs in $T$ such that its \pkt\
$x\in M\2$ is at $y\in T\2$.\ms

In the case that $\tl$ is one dimensional its \ti s may by ordered so that
$\tl$ may be understood as a two-sided sequence over $\Z$ having values in
the set of \tii s.
We simply write $\tl\2\cong\Z$ (isomorphic as ordered sets) in that case which
however does not mean that
$\tl$ is periodic. Correspondingly, for Cartesian products of one dimensional
tilings we would write $\tl\2\cong\Z^d$. Apart from these kind of tilings those
with nontrivial orientational symmetry will be considered, since they are of
particular importance as idealized models for the spatial structure of
quasicrystals.
\bs

\subsection{Discrete models of solids}

As already mentioned in the introduction, the gap labelling is a qualitative
characterization of the spectrum of a \So\ which incorporates the structure of
the underlying space.
The spatial structure of a solid may be described by a tiling. The atoms are
located in the center of the \ti s their type and distribution being encoded in
the local pattern surrounding them.
Since a tiling is to reflect only this spatial structure but not the details
as e.g.\ the bindings -- these would be taken care of in a specification
of the \So\ -- it should be expected that two tilings whose local patterns may
be
derived from each other by inspection of only finite patches lead to the same
results for the gap labelling.\ms

The \So s for  one or many particle problems are considered here in a discrete
version.
They are assumed to be local and to depend only on the \miii es not exceeding
a given finite size\footnote{
Actually, limits with respect to a certain $C^*$-norm could also be allowed.}.
In particular they are selfadjoint operators on $\ell^2(\tl\2)$ and
the above dependence on the \miii es more precisely means that the action of
such an operator $H$ with interaction radius $r$
is of the form (slightly simplifying the notation):
\be  \label{22pa1}
(H\psi)(x) = \sum_{x'\in M\2}H_{xx'}\psi({x'})
\ee
where $(M,x)\subset(T,x)$ is the largest \mi\ of $\tl$ at $x$ which is covered
by an $r$-ball around $x$, and $H_{xx'}\in\Complex$ depends only on the
\miii\ of $M$, $x$, and $x'$.
The simplest tight binding model
 is $H=-\Delta+V$ where $\Delta_{xx'}$ is nonzero
and equal to 1 if and only if the \ti\ at $x$ and the \ti\ at $x'$ have a
common hypersurface
%(i.e.\ a link in two and a vertex in one dimension)
and $V$ is diagonal
$V_{xx}$ only depending upon the \tiii\ which is represented at $x$.

\subsection{The gap labelling}

Let us briefly describe how to assign elements of a countable Abelian group
(the labels) which are
insensitive under certain kind of perturbations to the possible gaps of the
spectrum of a \So .
This may be motivated by Shubin's formula which formulates the integrated
density of states (IDS) as a trace per volume evaluated on
spectral projections of the operator; but this formula is not a necessary
prerequisite.
 A more detailed description of the gap labelling may be found in the works
of Bellissard \cite{Be1,Be2,BBG}.\bs

Let $H$ be a bounded selfadjoint operator on $\ell^2(\tl\2)$
of the kind described above and $\Ch(H\leq E)$ its spectral projection onto the
space spanned by
eigenfunctions of eigenvalues smaller or equal to $E$. Denote by $\Ch_\Lambda$
the orthogonal projection onto $\ell^2(\Lambda\2)$, the space of wavefunctions
vanishing outside a finite \mi\ $\Lambda$ of
  $\tl$ and set $H_\Lambda=\Ch_\Lambda H\Ch_\Lambda$.
Let $|\Lambda\2|$ denote the number of \ti s in $\Lambda$.
To define the notion of an IDS on infinite dimensional Hilbert spaces one makes
the ansatz
\be                                 \label{11}
{\cal N}_H(E) = \lim_{\Lambda\rightarrow \tl}
\frac{1}{|\Lambda\2|} \mbox{Tr}(\Ch(H_\Lambda\leq E))
\ee
the limit being defined by an increasing chain $\Lambda_n\subset\Lambda_{n+1}
\subset\cdots$ of \mi s of $\tl$ which approximate $\tl$ and Tr being the
usual operator trace on
${\cal B}(\ell^2(\tl\2))$; hence
$\mbox{Tr}(\Ch(H_\Lambda\leq E))$ equals the number of eigenstates of
$H_\Lambda$ to eigenvalues smaller or equal than $E$. This notion of an IDS
is very important in solid state physics and is at least in principal
accessible
by experiments, c.f.\ \cite{Be1}.\ms

Questions of existence of the above limit and of its
independence of the chain of \mi s approximating $\tl$ as well as of the
boundary
conditions chosen for the finite approximant $H_\Lambda$ (here Dirichlet
boundary conditions are used) have not yet been answered in full generality.
However at least for the case that $\tl\2$ may be identified with an amenable
group and $E$ lies in a gap may the r.h.s.\ of (\ref{11}) be understood as
image of a trace evaluated on $\Ch(H\leq E)$. More precisely does
\be                                 \label{12}
\tr(a) = \lim_{\Lambda\rightarrow \tl}
\frac{1}{|\Lambda\2|} \mbox{Tr}(\Ch_\Lambda a) ,
\ee
define in that case a normalized trace on a suitable subalgebra of
${\cal B}(\ell^2(\tl\2))$, which is proved by Birkhoff's ergodicity theorem,
and amenability implies \cite{Be1}
\be                                \label{140}
\lim_{\Lambda\rightarrow \tl}
\frac{1}{|\Lambda\2|} \mbox{Tr}
(\Ch(H_\Lambda\leq E)-\Ch_\Lambda\Ch(H\leq E)) = 0 .
\ee
A discrete group $G$ is amenable if there exists a sequence of bounded subsets
$\{\Lambda_n\}_{n\geq 1}$ approximating $G$, i.e.\ $\bigcup_{n\geq 1}
\Lambda_n = G$ such that  $\forall g\in G:\frac{
|(\Lambda_n\backslash g\Lambda_n)\cup(g\Lambda_n\backslash\Lambda_n)|}
{|\Lambda_n|} \stackrel{n\rightarrow\infty}{\longrightarrow} 0$. What is
important here is not so much the group structure but the fact that the
volume of the \saum\ of $\Lambda_n$, i.e.\ the number of elements which are
just at the \saum\ of $\Lambda_n$, is negligible compared to the total volume
of $\Lambda_n$ if $n$ increases. Therefore Bellissard's proof of (\ref{140})
carries over to any tiling of the kind considered further down.
(\ref{12}) together with (\ref{140}) yield Shubin's formula
\be                                 \label{15}
{\cal N}_H(E) = \tr(\Ch(H\leq E)) .
\ee

Clearly ${\cal N}_H(E)$ is a monotonically increasing function of $E$
which is constant on gaps.

The unital \CA\ $C(H)$ generated by the selfadjoint $H$ is
isomorphic to $C(\sigma(H))$, the complex continuous functions over
the spectrum of $H$. If $E$ lies in a gap, i.e.\ if $E\notin\sigma(H)$, then
the characteristic function
$\Ch_{\{\lambda\in\sigma(H)|\lambda\leq E\}}\in C(\sigma(H))$
and therefore $\Ch(H\leq E)=\Ch_{\{\lambda\in\sigma(H)|\lambda\leq E\}}
(H)\in C(H)$, whereas for $E\in\sigma(H)$ the spectral projection
$\Ch(H\leq E)$ may in general only be found in the von Neumann closure of
$C(H)$.
Now if $\A$ is \CA\ with a normalized trace $\tr$ having a representation $\pi$
on $\ell^2(\tl\2)$ such that $H=\pi(h)$ for some selfadjoint $h\in\A$ and
 Shubin's formula holds with that $\tr$, then for $E\notin\sigma(H)$
 \be                                                  \label{13}
{\cal N}_H(E) = \tr(\Ch_{\{\lambda\in\sigma(H)|\lambda\leq E\}}(h)) \in
\tr(Proj(\A))
\ee
as the trace of a projection depends only on its equivalence class.
Two projections are equivalent $p\sim q$ whenever $\exists u\in\A:p=uu^*,
q=u^*u$, and $Proj(\A)$ denotes the set of equivalence classes.
This beautiful result shows a remarkable property of the IDS:
the values of the IDS on gaps are stable under
perturbations of $H$ which are connected to $H$ by a continuous path in $\A$
(continuity is meant here in the norm resolvent sense, c.f.\ \cite{ReSi}).
This together with the countability  of $\tr(Proj(\A))$ for separable $\A$
follow from the fact that two projections $p,q$ which are close to each other
in the sense of $\|p-q\|<1$ are equivalent.
Hence the values of the IDS on gaps furnish a countable set of labels for the
gaps which is stable under certain
perturbations.\bs

To compute the possible values of the IDS on gaps, first, a suitable \CA\ has
to be constructed.
The set of its projection classes should be as small
 that at least most elements of $\tr(Proj(\A))$  will actually occur as labels
for gaps. Of course, $\A$ should be related to $H$, but since $Proj(\A)$ does
not refer to a specific operator it will only be possible to obtain
statements for "generic" operators.
Actually, it is not only the gap structure of a specific discrete
\So\ which is of
interest but the generic features of a whole family.
At least in simple models the spectrum of an operator over a space continuum
may be locally determined by a whole family of discrete operators which
act on the Hilbert spaces of the tight binding approximation \cite{Be1}.

Concerning now the choice of this algebra we take the point of view that it
should be constructed purely from the spatial structure of the underlying
tiling. In mathematical terms this is the \CA\ which is
assigned to the \nc\ space of the tiling, in
physical terms it is the algebra of observables namely it is generated by
translations and characteristic functions.
In \cite{Be1}, where tilings are considered which may be ordered like
$\Z^d$, a \CA\ is
constructed
starting from a generic \So\ and its translates. This approach is very
closely related to the one considered here -- e.g.\ the algebraic hull of
the \So\ defined in \cite{Be1} is homeomorphic to the geometric hull discussed
below -- and leads in the comparable cases to the same algebra.\bs

$Proj(\A)$ is in general difficult to compute as it does not carry enough
structure. If instead the stabilized algebra $M_{\infty}(\A)$ is used the
structure of an addition may be defined on its projection classes leading
via Grothendieck construction to an Abelian group, the $K_0$-group of $\A$,
c.f.\ the
appendix. This group is naturally ordered, the projection classes of
$M_{\infty}(\A)$ being the positive elements, and scaled, the class of the
unit $1\in\A$ serving as order unit.
The trace on $\A$ induces a state $\tr_*$ on that group, i.e.\ a positive
homomorphism into $\Real$ normalized to $\tr_*[1]=1$.
This allows for a formulation of (\ref{13}) which may be slightly weaker:
\be                                                  \label{14}
{\cal N}_H(E) \in \tr_*(K_0(\A))\cap[0,1] ,
\ee
if $E$  lies in a gap. This is part of the abstract gap labelling theorem of
Bellissard. In many interesting cases the r.h.s.\ may be computed yielding a
set
of possible gap labels. In some cases one can show by different methods --
being more and less rigorous -- that any element of $\tr_*(K_0(\A))\cap[0,1]$
actually occurs as a value of the IDS on a gap of a generic operator
\cite{Be3,BBG2,Luc}. The period doubling \sst\ is an example where this
agreement could be shown rigorously \cite{BBG2}.
In case $K_0^+(\A)$ is already given by
$\{z\in K_0(\A)|\tr_*(z)>0\}\cup\{0\}$ and $\A$ has cancellation (\ref{14}) is
in fact equivalent to (\ref{13}) namely
$\tr(Proj(\A))=\tr_*(K_0(\A))\cap[0,1]$.
\ms

As $\tr_*$ is a homomorphism of groups, $\tr_*(K_0(\A))$ is a subgroup of
$\Real$ which is countable, if $\A$ is separable. However it should be noted
that the elements of $K_0(\A)$ themselves yield a set of possible labels that
are stable under perturbations as described above and which are ordered. This
way of labelling the gaps is independent of the validity of the Shubin formula,
but without such a formula a gap label cannot be related to the location of the
 gap in the spectrum. On the other hand, if $\tr_*$ is not injective, the gap
labelling by elements of $K_0(\A)$ is finer as the one by values of the IDS.\bs

{\em Remark:}
Even if the
elements of $Proj(\A)$ would lead to a finer gap labelling than the ones of
$K_0(\A)$
the stabilization
of the algebra may be advantageous. In \cite{Be2} it is shown that for the case
$\tl\2\cong\Z^d$ the algebra assigned to $\tl$ has the same stabilization as
the one which would be constructed for the non discrete problem. In other
words,
the spectrum of a generic operator defined on the space continuum is as well
characterized by the elements of $K_0(\A)$ as its tight binding approximation.

\section{The non commutative geometry of a tiling}

\subsection{The non commutative space of a tiling}

For a tiling $T$ of the
Euclidean space $\Real^d$ with distinguished origin $0\in\Real^d$
let $M_r(T)$ denote the smallest \mi\ of $T$ covering $B_{r}(0)$, the $r$-ball
of $0$. $T$ is called locally finite if $M_r(T)$ contains for finite $r$
only finitely many tiles.
The Euclidean group of isometries,  the semi-direct product of $O(d)$
with the  translations, acts on the set $\Tl$ of all
locally finite tilings of $\Real^d$ in a natural way.
This action is denoted by $x\cdot T$ or,
if $x$ is a translation, by $T\!-\!x$.
Define the coincidence radius of two tilings by
$r(T,T') := \sup\{r'\geq 0| M_{r'}(T)=M_{r'}(T')\}$.
Then $r(T,T'')\geq \min\{r(T,T'),r(T',T'')\}$ for any third $T'$, and
\be
d(T,T') := \exp(-r(T,T'))
\ee
defines
a metric on $\Tl$ which is invariant under $O(d)$ but not under
translations.

$\Tl$ is complete: Let $\{T^{(i)}\}_{i\geq 1}$ be a Cauchy
sequence in $\Om$. Then $\forall r>0\exists N_r\forall n,m\geq N_r:
d(T^{(n)},T^{(m)})<\exp(-r)$ and hence $\forall n\geq N_r:
M_r(T^{(n)})=M_r(T^{(N_r)})$. The increasing chain of  \mi s
$M_r(T^{(N_r)})\subset M_{r'}(T^{(N_{r'})})\cdots$, $r<r'\cdots$
defines a tiling of $\Tl$ which is the limit of the Cauchy
sequence.

$\Tl$ is totally disconnected (zero dimensional):
The metric topology is generated by the open $\epsilon$-balls
\be
U_\epsilon(T)=\{T'\in\Tl|\,d(T,T') < \epsilon\}
=\{T'\in\Tl|\,r(T,T') > -\ln\epsilon\}.
\ee
Local finiteness of $T$ implies that the image of the function
$d(T,\cdot):\Tl\rightarrow [0,\infty), T'\mapsto d(T,T')$ is discrete
so that $U_\epsilon(T)$ is as well
a closed $\epsilon+\delta$-ball for sufficiently
small $\delta$. Hence the topology is generated by open and closed sets
which is the definition of a zero dimensional space and implies total
disconnectness.
\ms

Now consider a fixed tiling $\tl\in\Tl$ having
 the \pkt\ of one of its  \ti s sitting on
$0\in\Real^d$, i.e.\ $0\in\tl\2$.

\bd
The set
$$ \Om := \overline{\{\tl-x|x\in\tl\2\}} $$
is called the hull of $\tl$.
\ed
Eventually, we shall be more precise and
write $\Om_\tl$ for the hull of $\tl$.
Clearly $\Om$ is finite if and only if $\tl$ is periodic in $d$ independent
directions. As $\Om$ is the closure
of a subset of  $\Tl$, it is complete.  By construction any
element $T$ of $\Om$ can be approximated by translates of $\tl$,
i.e.:
\be                \label{1ack4} \forall \epsilon>0\exists
x\in\Real^d: d(\tl\!-\!x,T)<\epsilon .
\ee
As a consequence, for any
  $T\in\Om_\tl$ and any \mi\ $(M,x)\subset (T,y)$
there is a $y'\in {\tl}\2$ such that $(M,x)\subset (\tl,y')$.  In
this sense $T$ cannot be distinguished from $\tl$ by inspection of its finite
patterns only and is consequently called {\em locally homomorphic} to $\tl$.
If $T\in\Om_\tl$ then also
$T-x\in\Om_\tl$ for all $x\in T\2$. Completeness of $\Om_\tl$ therefore
implies that $\Om_T\subset \Om_\tl$.
If the converse is true as well, i.e.\ if $\forall (M,x)\subset (\tl,y)\exists
y'\in T\2 :(M,x)\subset (T,y')$, then $T$ and $\tl$ are
locally isomorphic and $T$ defines
the same hull as $\tl$. We call a tiling $\tl$ {\em homogeneous}, if it is
locally
isomorphic to any element of its hull.

The relative topology on $\Om$ is generated by the sets
$U_{\exp(-r)}(T)\cap\Om$ where we may restrict to the values of
$r$ which do not lie in the image of $r(T,\cdot)$ and for which
\be       \label{2102}
U_{\exp(-r)}(T)\cap\Om=U_{M_r(T),0}:=\{T'\in\Omega|\, M_r(T) = M_r(T')\}.
\ee
More generally, the following sets which are indexed by \mii es and one of
their \pkt s
\be
U_{M,x}:=\{T\in\Om|(M,x)\subset (T,0) \}
\ee
are open and closed, they will be of importance later on.
It is clear that either $U_{M_r(T),0}=U_{M_r(T'),0}$ -- namely if
$M_r(T)=M_r(T')$ -- or $U_{M_r(T),0}\cap U_{M_r(T'),0}=\emptyset$.
We shall restrict ourself to tilings which lead to hulls that satisfy the
following condition.
\bi
\item[\bf \bo]
For any $r\geq 0$ there are only finitely many tilings
$T_i\in\Omega$ such that $M_r(T_i)$ are pairwise distinct.
\ei
This implies that $\tl$ has only finitely many \tii s.
\bl                                     \label{l1}
If $\Om$ satisfies \bo\ it is compact.
\el
{\em Proof:}
According to condition \bo, to a  given $r\geq 0$
there is a maximal finite family $\{T_i\}_{i\in I}$  of tilings
such that  $M_r(T_i)$ are pairwise distinct. Hence
\be         \label{2104}
\Omega  = \bigcup_{i\in I} U_{M_r(T_i),0} =
\bigcup_{i\in I}\{T'\in\Omega|\,d(T_i,T') < \exp(r_0-r)\}
\ee
for some $0\leq r_0 < r$ so that
$\Om$ has a finite cover of open $\exp(r_0-r)$ balls.
By varying $r$ and $r_0$ the number $\exp(r_0-r)$ may take any small positive
value so that $\Om$ is precompact which for  complete spaces is equivalent to
compactness.
\eb \bs

Although this topology might at first sight appear a little bit unusual, it has
a physical meaning in that homeomorphic hulls are defined by locally
equivalent tilings: Let $\tl_i$, $i=1,2$ be two
tilings satisfying \bo, write $\Om_i^0=\{\tl_i-x|x\in\tl_i\2\}$,
i.e.\ $\Om_{\tl_i}=\overline{\Om_i^0}$, and let
$\jk:\tl_1\2\rightarrow \tl_2\2$ be a map. $\jk$ then defines
a mapping
$\jk:\Om^0_1\rightarrow\Om^0_2$
\be
\jk(\tl_1-x) := \tl_2-\jk(x)
\ee
which extends to the closures, if it is continuous. But continuity at
 $T\in\Om^0_1$ means that
\be
\forall r_2>0\exists r_1>0: \jk(U_{M_{r_1}(T),0}\cap\Om^0_1)
\subset U_{M_{r_2}(\jk(T)),\jk(0)}\cap\Om^0_2 ,
\ee
i.e.\ the \mi\ around $0$ of $\jk(T)$ covering $B_{r_2}(\jk(0))$ is already
determined by the \mi\ of $T$ which covers $B_{r_1}(0)$.
One is used to
say that $\jk(T)$ is locally derivable from $T$. If moreover $\jk$ is bijective
and therefore a homeomorphism,
then $T$ and $\jk(T)$ are called {\em locally equivalent}. The spatial
structure described by them should then lead to the same physics \cite{BKS}
which is
reflected here by the fact that they have homeomorphic hulls.

A special case of local equivalence arrises if one has two tilings
which differ only by the \pkt\ of their \tiii s,
as long as they are both generic.
We call a puncture generic if it maximally breaks the symmetry of the
\tiii.
Changing a generic \pkt\ to a non generic one may result in a change
of the hull as a topological space. Namely if a \pkt\ $x\in M\2$ lies on a
symmetry axis or  centre of some symmetry $g\in O(d)$ of the
\mi\ $M$ then  $U_{g\cdot M,g\cdot x}=U_{M,x}$, whereas otherwise
 $U_{g\cdot M,g\cdot x}\cap
U_{M,x}=\emptyset$.
\bl The hulls defined by tilings which
differ only in their \pkt\ are homeomorphic, if both \pkt s are
generic. \el
{\em Proof:}
Let $\tl_1$, $\tl_2$ denote the tiling with the first resp.\
the second \pkt\ and $\jk:\tl_1\2\rightarrow\tl_2\2$ be the map assigning
the first \pkt\ of a \ti\ to the second. Since the \pkt s are generic to any
 \mii\ $M_1$ of $\tl_1$ corresponds
a unique \mii\ $M_2$ of $\tl_2$ which differs only by the \pkt. Therefore
$\jk(U_{M_1,x}\cap\Om^0_1) = U_{M_2,\jk(x)}\cap\Om^0_2$ which shows that $\jk$
extends to a homeomorphism.\eb
\bs

Punctures are always assumed to be generic.\ms

Everything that will be treated in this article depends only on the
topological structure of $\Om$. So we might in principle talk about
the \nc\ space resp.\ the algebra of a local equivalence class of a tiling,
but we prefer the simpler expressions.

The action of the subgroup of translations restricted to $\Om$
yields an equivalence relation which we denote by
$\Gr=\{(T,T')\in\Om\times\Om|\exists
x\in\Real^d:T'=T\!-\!x\}$.
\bd The non commutative space of the tiling
$\tl$ is the quotient
$$ \Psi := \Om/\!\sim_\Gr .$$
\ed
In other words,
$\Psi$ is the space of all tilings which are locally
homomorphic to $\tl$ modulo what may be called globally isomorphic ones.
If $\tl$ is periodic in $d$ independent directions then $\Psi$ is one point.
However otherwise it is non
Hausdorff with respect to the quotient topology, as any tiling $T$ is arbitrary
close to some translate of $\tl$, cf.\ (\ref{1ack4}).
In a sense $\Psi$ is then a point
 with an additional
structure which is undetectable by commutative geometry
\cite{Con}. A.\ Connes' proposal to analyse such spaces leads via
the construction of non commutative \CA s to the study of their
invariants, e.g.\ their \kg s. This will be the guiding line below. \bs

A subclass of aperiodic tilings which are of particular interest
for us are those which have an orientational symmetry
\cite{Soc}. An element $g\in O(d)$ is called an {\em orientational
symmetry} of $\tl$ if $\forall r\geq 0\exists
x\in\tl\2:M_r(g\cdot\tl)=M_r(\tl\!-\!x)$. An element $T\in\Om$ is called
symmetric under $g\in O(d)$ if $\exists x\in T\2 : g\cdot T = T\!-\!x$.
The orientational
symmetries yield a subgroup of $O(d)$ which we shall denote by
$\Sy$. If $\tl$ satisfies condition \bo\ $\Sy$
is a finite group.
If $T\in\Om$ then also
$g\cdot T\in\Om$, provided $g\in\Sy$. Hence $\Sy$ acts on $\Om$,
and this action is continuous as $O(d)$ preserves the metric.\bs

{\em Remark:} The notion of
the local isomorphism class of a tiling has been introduced for the
description of quasicrystals.
A subclass thereof, the closed subset
$$ \LI =\{T\in \Tl|\forall r\geq 0\, \exists x,y\in \Real^d: M_r(\tl\!-\!x)
= M_r(T)\:\mbox{ and }\: M_r(\tl) = M_r(T\!-\!y) \}$$ of $\Tl$ is
closely related to the hull of $\tl$. It may be shown that any
element of the quotient space $\LI/\mbox{\small transl.}$ contains a
representative in $\Om_\tl$.

\subsection{The algebra of a tiling}

We now define a \CA\ $\A_\tl$ to a given tiling $\tl$. As
already mentioned this algebra will be constructed from the
\nc\ space $\Psi_\tl=\Om_\tl/\!\sim_\Gr$, namely it will be the
groupoid-\CA\ of the equivalence relation. From a physicists
point of view it is the algebra of observables. This algebra has
been introduced in a similar context by J.\ Bellissard
\cite{Be1,Be2}.
It yields an application of the theory of groupoid $C^*$-algebras
developed by Renault \cite{Ren}
which can be carried through once a suitable topology for the groupoid
has been found. \bs

A topological groupoid $\Gru$ is a topological space together
with a continuous product map $(x,y)\mapsto xy$:
$\Gru^2\rightarrow\Gru$ which is however only defined on a
subset  $\Gru^2\subset\Gru\times\Gru$ and a continuous inversion
map $x\mapsto x^{-1}$: $\Gru\rightarrow\Gru$, such that the
following relations hold:
\begin{enumerate}
\item $(x^{-1})^{-1} = x$,
\item if $(x,y),(y,z)\in\Gru^2$ then $(xy,z),(x,yz)\in\Gru^2$ and
 $(xy)z = x(yz)$,
\item $(x^{-1},x),(x,x^{-1})\in\Gru^2$; moreover if
$(x,y)\in\Gru^2$ resp.\ $(z,x)\in\Gru^2$, then $x^{-1}(xy) = y$
resp.\ $(zx)x^{-1} = z$.
\end{enumerate}
Consequently the elements of $\Gru^0=\{x^{-1}x|x\in\Gru\}$ are
called units. \ms

Like any equivalence relation  $\Gr$ is
a groupoid, namely $\Gr^2=\{((T,T'),(T',T'')) | T\sim T'\sim
T''\in\Om\}$ and the product is given by transitivity,
$(T,T')(T',T'') = (T,T'')$, and inversion by reflexivity
$(T,T')^{-1}=(T',T)$. Units are elements of the form $(T,T)$,
hence $\Gr^0\cong\Om$. The topology of $\Gr$ is defined to be the one
generated by the sets
\be
U_{M,x,x'} :=\{(T,T-\!(x'\!-\!x))|\,T\in U_{M,x}\}
\ee
where $M$ is an \mii\ of $\tl$ and $x,x'\in M\2$.
In this topology  $\Gr$ is locally compact,
$\Gr^0$ open and $\Gr^T=\{(T,T-x)|x\in T\2\}$ discrete. Thus in the teminology
used in \cite{Ren} $\Gr$ is a principal $r$-discrete groupoid. Note that the
above topology does not coincide with the relative topology inherited from
$\Om\times\Om$!

On the space of all continuous complex functions with compact
support $C_c(\Gr)$ one introduces a product and an involution by
\begin{eqnarray}                             \label{223}
f*g\,(T,T') & = & \sum_{T'' \sim T} f(T,T'')\: g(T'',T') , \\
f^*\,(T,T') & = & \overline{f(T',T)}        \label{223b}
\end{eqnarray}
thus obtaining a topological  $*$-algebra. Restriction to
functions with compact support leads to a sum in (\ref{223})
which contains only finitely many nonzero terms.
$C_c(\Gr)$ is generated as a $*$-algebra by the functions
$e_{M,x,x'} : \Gr\rightarrow \Complex$
\be e_{M,x,x'} (T,T') :=
\left\{ \begin{array}{l} 1\: \hbox{ if }\:(M,x)\subset
(T,0)\:\hbox{ and }\: T'= T-(x'\!-\!x) \\ 0\: \hbox{ else}
\end{array}      \right.
\ee
which correspond to the
characteristic functions on  $U_{M,x,x'}$.

The \CA\ $\A_\tl$ associated to the tiling $\tl$ is defined
to be the completion of $C_c(\Gr)$ with respect to the reduced
norm. For the definition of this norm one considers the family
of $*$-representations of  $C_c(\Gr)$, which we call physical
representations, and which are labelled by a tiling
$T\in\Omega$: $\pi_T$ acts on the Hilbert space $\ell^2(\Gr^T)$
via
\be                                           \label{225}
(\pi_T(f)\psi)(T') := \sum_{T''\sim T'} f(T',T'')\psi(T'') ,
\ee
$\psi\in\ell^2(\Gr^T)$ and writing shorter $\psi(T')$ for
$\psi(T,T')$.\footnote{
It is more convenient here to formulate this representation for wavefunctions
over $\Gr^T$ the connection to those on $T\2$ being obvious.}
 In such  a
representation $e_{M,x,x'}$ becomes  translation by $x'-x$ in the following
sense:
\be                            \label{224}
(\pi_T(e_{M,x,x'})\psi)(T') = \left\{ \begin{array}{ll}
\psi(T'-(x'\!-\!x)) & \hbox{if }\:(M,x)\subset (T',0) \\ 0 &
\hbox{else} \end{array}      \right. .
\ee
These representations are irreducible, in fact the commutant of
$\pi_T(C_c(\Gr))$ in $\B(\ell^2(\Gr^T)$ may be seen to be trivial.
The reduced norm of
an element of  $C_c(\Gr)$ is the supremum of its operator norms
in these representations, $\|f\|_{red}=\sup_{T}\|\pi_T(f)\|$,
and the reduced groupoid-\CA\ $C^*_{red}(\Gr)$ is the completion
of  $C_c(\Gr)$ with respect to this norm.
$\A_\tl=C^*_{red}(\Gr)$ may be considered as the algebra of
observables of a quantum mechanical system, as it is generated
by translations. \bs

{\em Remark:}  The product on $C^*_{red}(\Gr)$
which looks like matrix multiplication is in fact a convolution.
To define it in the
general case one has to introduce a Haar system on $\Gru$.
A left-Haar system on $\Gru$ is a family of measures
$\{\lambda^u,u\in\Gru^0\}$ on $\Gru$ that satisfy
\begin{enumerate}
\item the support of  $\lambda^u$ is $\Gru^u = \{x\in\Gru|x
x^{-1} = u\}$, \item for every $f\in C_c(\Gru)$ the map
$u\mapsto \int_{\Gru} f(x)\,d\lambda^u(x)$ is continuous, \item
(left-invariance) $\int_{\Gru} f(xy)\,d\lambda^{x^{-1}x}(y) =
\int_{\Gru} f(y)\,d\lambda^{xx^{-1}}(y)$ for all $x\in\Gru$ and
$f\in C_c(\Gru)$.
\end{enumerate}
A convolution and an involution may then be defined on
$C_c(\Gru)$ by
\begin{eqnarray}                            \label{221} f*g\,(x)
& =&  \int_{\Gru} f(xy)\, g(y^{-1})\,d\lambda^{x^{-1}x}(y)\\
f^*\,(x) & =&  \overline{f(x^{-1})}         \label{222}
\end{eqnarray}
In our case $\Gr^{(T,T)}=\Gr^T$ is discrete so that left invariance forces the
measure $\lambda^T=\lambda^{(T,T)}$ on $\Gr^T$, if it is
normalized to $\lambda^{(T,T)}(T,T) = 1$, to be the point
measure \cite{Ren}, i.e.\ $\lambda^T(T',T'')=\delta_{TT'}$. Then
(\ref{221},\ref{222}) become (\ref{223},\ref{223b}).

\subsubsection*{A trace on $\A_\tl$}

With the help of an $\Gr$-invariant measure on $\Om$ a trace on
$\A_\tl$ can be defined.
A measure $\mu$ on $\Om$ is called invariant (or more precisely
$\Gr$-invariant)
if $\mu(U_{M,x})$ is independent of the choice of $x\in M\2$.
Let $\mbox{P}:\,C^*_{red}(\Gr)\rightarrow C(\Omega)$ be the
restriction map. $\mbox{P}$ is a generalized conditional
expectation, and provided $\mu$ is invariant
\be
\tr(f):=\int_\Omega\mbox{P}(f)\,d\mu
\ee
defines a trace on $C^*_{red}(\Gr)$. Its  cyclicity is a direct consequence
of the invariance of $\mu$, as
\be
\tr(f*g) =
\int_\Omega (f*g)(T,T)\,d\mu(T) =
\int_\Omega \sum_{T'\sim T}f(T,T')\,g(T',T)\,d\mu(T) .
\ee
Concerning the gap-labelling we are of course interested in
traces that coincide with the one
appearing in the r.h.s.\ of the Shubin formula.
If $\tl\2\cong\Z$ and $\mu$ is ergodic, then one can
show by  Birkhoff's ergodic theorem that the r.h.s.\ of
(\ref{12}) converges for $\mu$-a.e.\ $T$ to the trace defined by
$\mu$ \cite{Be1}. Together with (\ref{140}) this implies the
validity of the Shubin formula. In the general case we merely
can establish the following: In a representation
of the kind (\ref{225}) $\mbox{P}$ is the projection onto the
diagonal and therefore  $\mbox{Tr}\circ \mbox{P} = \mbox{Tr}$.
This implies for arbitrary $a\in\A_\tl$
\be                \label{233} \lim_{\Lambda\rightarrow \tl}
\frac{1}{|\Lambda\2|} \mbox{Tr}(\Ch_\Lambda \pi_T(a)) =
\lim_{\Lambda\rightarrow \tl} \frac{1}{|\Lambda\2|}
\mbox{Tr}(\Ch_\Lambda \pi_T(\mbox{P}(a)))
\ee
and hence
(\ref{12}) holds for arbitrary $a\in\A_\tl$, if for all $f\in
C(\Omega)$
\be                                 \label{234}
\lim_{\Lambda\rightarrow \tl} \frac{1}{|\Lambda\2|}
\mbox{Tr}(\Ch_\Lambda \pi_T(f)) =  \mu(f) := \int_\Om f \,d\mu.
\ee
We shall
construct measures satisfying this equality. \bs

The $K_0$-group of $\A_\tl$ is not known to us in the general
case neither the complete image of the tracial state. But such a
state induced by an invariant measure $\mu$ satisfies always
\be        \label{232}
\mu(C(\Omega,\Integer))  \subset\tr_*(K_0(C(\Om))\subset\tr_*(K_0(\A_\tl))
\ee
$C(\Omega,\Integer)$ denoting the continuous functions over
$\Om$ with values in $\Z$.
Hence part of the values of the
tracial state is already been given by the image of an invariant
measure on $\Om$, the normalization of which is determined by
the order unit: $\mu(\Om)=\tr_*[1]=1$.
However we emphasize that in
all cases known to us where $\tr_*(K_0(\A_\tl))$ can be computed
even equality holds in  (\ref{232}) which is not obvious from
the $K_0$-group.\bs

{\em Remark:}
For completeness let us briefly describe the notion of
$\Gr$-invariance in the general case \cite{Ren}.
Let $\mu$ be a measure on $\Gru^0$ and
$\{\lambda^u|u\in\Gru^0\}$ a left Haar system and set
$\lambda_u(x) =\lambda^u(x^{-1})$. Then $\nu =
\int_{\Gru^0}\lambda^u\,d\mu(u)$ as well as $\nu^{-1} =
\int_{\Gru^0}\lambda_u\,d\mu(u)$ are in general two different
measures on $\Gru$. $\mu$ is called invariant, if for the Radon-Nikodym
derivative
$\frac{d\nu}{d\nu^{-1}}=1$.
Applied to our case this one obtains
\be         \label{4pa1a}
\nu(U_{M,x,x'}) =
\int_{\Omega}d\mu(T)\int_{U_{M,x}}d\lambda^T(T',T'-\!(x'\!-\!x)) =
\mu(U_{M,x})
\ee
as well as
\be       \label{4pa1b}
\nu^{-1}(U_{M,x,x'}) =
\int_{\Omega}d\mu(T)\int_{U_{M,x}}d\lambda^T(T'-\!(x'\!-\!x),T')
 = \mu(U_{M,x'})
\ee
leading to the description of invariance of above.
The conditional  expectation satisfies
\be      \label{231}
\int_\Omega \mbox{P}(g^**f)\,d\mu =\int_\Gr f\overline{g}\,d\nu^{-1}
\ee
which coincides with the scalar product $<\!f,g\!>$ of elements in
$L^2(\Gr,\nu^{-1})$ so that the latter can also be seen as
$L^2(\overline{C^*_{red}(\Gr)},\tr)$ where $\overline{C^*_{red}(\Gr)}$ is
 the completion of $C^*_{red}(\Gr)$ with respect to the scalar product defined
by the trace.

\section{Cartesian products of one dimensional
 tilings -- general considerations} \label{sec3}

In this section a general analysis for the gap labelling
of the simplest class of tilings is carried out.
The tilings of this class may be understood as mappings from $\Z^d$ into the
set of \tii s, i.e.\
$\tl\2\cong\Z^d$.
The results obtained are not yet very explicit and extra structure is needed
to compute the gap labelling.

\subsection{One dimensional tilings}

A one dimensional tiling in our sense is a covering of $\Real$ by \pkt d
intervals and is therefore ordered.
A \tiii\ is determined by the length of the interval and the
position of its \pkt.
As usual the \tii s
are denoted by letters of a (finite) alphabet $\B =\{a_1,\cdots,a_n\}$.
A tiling is therefore a two-sided sequence of letters
$T = \{T_i\}_{i\in\Integer}$ and a \miii\ of $T$ is a word in the alphabet
which appears in $T$. We shall fix the letter whose \pkt\ sits on
$0\in\Real$ to be $T_0$. \ms

Also $T\2$ may be ordered and identified with $\Z$, so that the equivalence
relation may be expressed with the help of a
  homeomorphism
$\varphi$ which acts on a sequence by shifting it to the left,
$(\varphi T)_n = T_{n+1}$, i.e.\
\be          \label{8pa1}
\Gr=\{(T,\varphi^{-k}T)\in\Om\times\Om |k\in\Z\} .
\ee
Moreover, via identification $(T,\varphi^{-k}T)\cong (T,k)$ one obtains
$\Gr \cong \Omega\times \Ga$
the topology being the product topology, where $\Ga=\Integer_h$ if $\tl$ is
$h$-periodic,
and $\Ga=\Z$ if it is non periodic. Through the
above identification  (\ref{223}) becomes
\begin{equation}                             \label{311}
f*g\:(T,k)  = \sum_{m\in\Ga} f(T,m)\,
g(\varphi^{-m}T,k-m) .
\end{equation}
Equivalently, the functions
$\hat{f}:\Ga\longrightarrow C(\Omega)$
where $\hat{f}(k)(T) = f(T,k)$ yield the algebra
 $\A_\tl$ the product being
\begin{equation}                             \label{312}
\hat{f}*\hat{g}\:(k)  = \sum_{m\in\Ga} \hat{f}(m)\,
\varphi^*_m\hat{g}(k-m)
\end{equation}
the involution
$\hat{f}^*\,(k) = \overline{\varphi^*_k\hat{f}(-k)}$. Here
$\varphi^*_m\hat{f} = \hat{f}\circ{\varphi}^{-m}$ is the pullback action of
$\Ga$ on $C(\Omega)$. Summarizing $\A_\tl\cong C(\Omega)\times_{\varphi}\Ga$
is a crossed product. (In particular, the reduced norm coincides with
the usual one, since $\Ga$ is amenable.)
 It may also be obtained as the $C^*$-hull of the $*$-algebra which is
generated
by $C(\Omega)$ together with
the function
$u:\Ga\rightarrow C(\Omega)$, $u(n) = \delta_{1n}$
 -- with $h$-periodic delta symbol for $\Ga=\Z_h$.
$u$ is unitary,  $u^{-1}(n) = \delta_{-1n}$, and generates the action of
$\varphi$ by conjugation: $u*\hat{f}*u^{-1} = \varphi^*\hat{f}$.
It is given by
\be
u = \sum_{a_ia_j\:\mbox{\scriptsize appears in}\:\tl}\hat{e}_{a_ia_j,1,2}
\ee
$1$ resp.\ $2$ abbreviating the \pkt\ of the first resp.\ second letter
in $a_ia_j$. In a representation $u$ will be represented as a translation
operator and $u+u^*$ as the discrete Laplacian.\ms

The pair
$(\Om,\varphi)$ is a topological dynamical system which is topologically
transitive as the orbit of $\tl$ is dense. If $\tl$
is homogeneous, i.e.\ $\forall T\in\Om_\tl:\Om_T=\Om_\tl$,
then any orbit is dense in which case the dynamical system is called
minimal. In the latter case $\A_\tl$ is a simple \CA.

\subsubsection*{Periodic tilings}

Let $\tl$ be a tiling of period $h$, i.e.\ it is
given by periodic repetition of
a word $a_1\cdots a_h$.
Then the shift action $\varphi$ is transitive and $\varphi^h\tl=\tl$.
Hence  $\Om$ consists of one orbit of $h$ elements, $\Om\cong\Gr^T$, and
$\A_\tl=\Complex^h\times_{\varphi}\Integer_h\cong M_h(\Complex)$.
Next to $K_1(M_h(\Complex))=\{0\}$ one obtains
\begin{eqnarray}
K_0(M_h(\Complex))  & = & \Integer \\
tr_*K_0(M_h(\Complex)) & = & \frac{1}{h}\Integer
\end{eqnarray}
for the $K_0$-group and the values of the state induced by the unique
(normalized) trace on $M_h(\Complex)$.
The representations on $\ell^2(\Gr^T)\cong \Complex^h$ are for different
$T$ unitary equivalent, and
$e_{a_ha_1,1,2} = e_{a_1\cdots a_h,h,1}$. Hence
$u = \sum_{i=1}^h\hat{e}_{a_ia_{i+1},1,2}$, $a_{h+1}=a_1$, is the translation
operator in this
representation which is a representation on the closed chain with $h$ sites
being obtained from $\tl$ by periodic identification.
$\A_\tl$ is the algebra of observables on this chain.\ms

If one would like to construct an algebra which has representations in which
a translation operator on the tiling itself is represented, one has to choose
in place of $u:\Integer_h\rightarrow C(\Omega)$ an operator the $h$-th
power of which is not equal to the identity operator.
This is guaranteed if one uses instead the \CA\
$C(\Omega)\times_{\varphi}\Integer$ (with the same $\varphi$),
namely the operator $\tilde{u}:\Integer\rightarrow C(\Omega)$,
$\tilde{u}(n) = \delta_{1n}$ corresponds to the translation operator on $\tl$
the $h$-th power of it lying in the centre of the algebra.
Now
\be                       \label{8pa2}
C(\Omega)\times_{\varphi}\Integer \cong
M_h(C(S^1))
\ee
coincides with the result following from Bloch's theorem for one dimensional
periodic models over $\Z$: As the \So\ does only depend on the local \mii es
it commutes with translation over $h$ lattice sites.
Therefore the Hilbert space of a representation is decomposed into the
eigenspaces of the $h$'th power of the translation operator. These eigenspaces
are $h$ dimensional and contain wavefunctions which are periodic up to a phase
which is the eigenvalue of the eigenspace. If this phase is absorbed
into the \So\ the operator can be understood as an $h\times h$ matrix having
coefficients depending on $S^1$.\ms

 The $K_0$-group of (\ref{8pa2}) is simply
\be
K_0(M_h(C(S^1)))  =  \Integer .
\ee
The choice of a normalized trace on $M_h(C(S^1))$ amounts to the choice of a
normalized measure $\mu$ on $S^1$,
\be
\tr(a) = \frac{1}{h}\int_{S^1} \mbox{Tr}(a(z))\,d\mu(z) .
\ee
But for projections $p\in M_h(C(S^1))$, i.e.\ $p(z)\in M_h(\Complex)$, is
$\mbox{Tr}(p(z))$ constant so that for all choices
\be
tr_*K_0(M_h(C(S^1)))  =  \frac{1}{h}\Integer .
\ee
As a result the gap labelling of periodic systems in one dimension is
independent
of whether one periodically identifies the tiling or not.
The difference shows up in the  $K_1$-group. This group is generated by the
class of $\tilde{u}$:
\be
K_1(M_h(C(S^1)))  =  \Integer .
\ee

\subsubsection*{Non periodic tilings}

Focussing now on non periodic tilings
$\Ga$ becomes  $\Z$. As the action of $\varphi$ on $\Om$ is topologically
transitive the \kg s of $\A_\tl$ may be computed with the help of the exact
Pimsner-Voiculescu sequence \cite{PiVo}. This has been carried out in
\cite{BBG,Put}.
\bt                               \label{thm31}
Let $\Omega$ be a  totally disconnected compact metric space, $\varphi$ a
topologically transitive homeomorphism on $\Om$ and $\tr$ a normalized trace
on  $C(\Omega)\times_{\varphi}\Integer$. Then
\begin{enumerate}
\item $K_0(C(\Omega)\times_{\varphi}\Integer) \cong
C(\Omega,\Integer)/E_{\varphi}$ where
$E_{\varphi} := \{f-f\circ \varphi^{-1}|f\in C(\Omega,\Integer)\}$
\item $K_1(C(\Omega)\times_{\varphi}\Integer) \cong\Integer$   its
generator being $[u]$.
\item $tr_*(K_0(C(\Omega)\times_{\varphi}\Integer)
 = \mu(C(\Omega,\Integer))$ where $\mu$ is the $\varphi$-invariant measure on
$\Om$ obtained by restriction of $\tr$ to $C(\Omega)$.
\item
$K_i(C(\Om)\times_\varphi\Z)$ has no nilpotent elements.
\end{enumerate}
\et

{\em Proof:} Proofs of the first three statements are given in
 \cite{BBG} or \cite{Put} so we restrict here to the last one.

For $K_1(C(\Om)\times_\varphi\Z)$ the statement is clear.
Let $n[f]=[0]$, $0\neq n\in \Z$, $f\in C(\Om,\Z)$ and $[f]$ being its
equivalence
class. Hence $\exists h\in C(\Om,\Z):\,nf=h-h\circ\varphi^{-1}$. Let
$\tilde{h}:=\frac{1}{n}(h-h(\omega))$ for some $\omega\in\Om$. Then
$f=\tilde{h}-\tilde{h}\circ\varphi^{-1}$, and since $\tilde{h}(\omega)=0$
which implies
$\tilde{h}(\varphi^{-n-1}(\omega))=\tilde{h}(\varphi^{-n}(\omega))
-f(\varphi^{-n}(\omega))\in\Z$ the function $\tilde{h}$ takes integer values
on the orbit of $\omega$. By the transitivity of $\varphi$ we may choose
$\omega$ in a way that its orbit is dense in $\Om$. Continuity of $\tilde{h}$
implies then $\tilde{h}\in C(\Om,\Z)$ and hence $f\in E_\varphi$, i.e.\
$[f]=[0]$. \eb
\bs

It should be noted that this theorem also applies to the algebra (\ref{8pa2}).
In the theorem a $\varphi$-invariant measure on $\Om$ was determined by
restricting the
trace to $f\in C(\Om)$: $\mu(f)=\int_\Om fd\mu=\tr(f)$.
Conversely any $\varphi$-invariant measure on $\Om$
defines a trace on the crossed product by $\tr(\hat{f}) = \int\hat{f}(0)d\mu$.
As there are no periodic orbits this is the only trace restricting to the
measure
$\mu$ \cite{Tom} and there is a one to one correspondence between
$\varphi$-invariant measures on $\Om$ and traces on $\A_\tl$.
Concerning the gap-labelling it is therefore important to know, under which
conditions such a trace coincides with the one of the r.h.s.\ of the
Shubin formula. Using Birkhoff's ergodicity theorem one can show that
this is the case if the corresponding measure is normalized and ergodic
\cite{BBG}. In case $\A_\tl$ has a unique normalized trace
the corresponding measure is ergodic.
Hence the values of the IDS on gaps are determined by an ergodic probability
measure on $\Om$. To explicitly determine such a measure or the
group $C(\Omega,\Integer)/E_{\varphi}$ an additional structure is necessary.
This additional structure will later on be the self-similarity of a
tiling. However before coming to that the Cartesian product case is discussed.

\subsection{Cartesian products of one dimensional tilings}

Among the simplest higher dimensional tilings are those which may be regarded
as
Cartesian products of one dimensional tilings. Their basic feature is that
$\tl\2$ may still be ordered as $\Integer^d$,
$d$ being the dimension of the tiling. By use of deep theorems from $K$-theory
the
\GL\ of this case may be traced back to the one dimensional case
and, if the values of the IDS are used as labels, it is again determined by
an invariant measure on the hull. We shall restrict here to the non periodic
case.
In fact, the periodic one may be treated in the same way if one takes
(\ref{8pa1}) as the relevant algebra.
\bs

First observe that through the identification of $\tl\2$ with $\Integer^d$
the groupoid-\CA\ $\A_\tl$ becomes in the non periodic case a crossed product
with $\Z^d$:
\be
\A_\tl = C(\Om)\times_\varphi\Z^d
\ee
where $\Om=\Om_1\times\cdots\times\Om_d$ is the Cartesian product of the one
dimensional hulls and $\varphi = \varphi_1\times\cdots\times\varphi_d$ is the
homeomorphism whose pullback gives the action of $\Z^d$.
\bl
Let $\tl_k$, $k=1,\cdots, d$ be one dimensional non periodic tilings. Then
\be                                    \label{601}
\A_{\tl}\cong\AT{1}\otimes \AT{2}\otimes\cdots \otimes\AT{d}
\ee
\el
{\em Proof:} As crossed products of nuclear \CA s with $\Z^d$ are nuclear
we only have to show that the algebraic tensor product
$C(\Omega_1)\times_{\varphi_1}\Z \:\odot\cdots\odot
C(\Omega_d)\times_{\varphi_d}\Z$ is $*$-isomorphic to a norm-dense subalgebra
of
$C(\Omega)\times_\varphi\Z^d$. In fact, the $*$-homomorphism
$a_1\odot \cdots\odot a_d\mapsto
a_1*\cdots* a_d$
from the algebraic tensor product onto the
crossed product with $\Z^d$ is injective. Moreover it has a dense image, since
$C(\Om)\cong C(\Om_1)\otimes \cdots \otimes C(\Om_d)$ and therefore
$C(\Omega)\times_\varphi\Z^d$
is generated as a \CA\ by elements of the form
$f_1\,\delta_{n_1}*\cdots *f_d\, \delta_{n_d}$, $f_k\in C(\Om_k)$.\eb
\bs

For the computation of the $K$-groups of $\A_\tl$ we may just apply the
K\"unneth formula (\ref{ab1},\ref{ab2}).
If $\tl$, $\tl'$ are two tilings which are Cartesian products of one
dimensional
ones this formula yields:
\begin{eqnarray}                \label{602a}
K_0(\A_{\tl\times\tl'}) &\cong&
K_0(\A_\tl)\otimes K_0(\A_{\tl'})\oplus
K_1(\A_\tl)\otimes K_1(\A_{\tl'})\\
 \label{602b}
K_1(\A_{\tl\times\tl'})& \cong& K_0(\A_\tl)\otimes K_1(\A_{\tl'})\oplus
K_1(\A_\tl)\otimes K_0(\A_{\tl'}) .
\end{eqnarray}
In fact as the $K$-groups of
$C(\Omega_k)\times_{\varphi_k}\Z$
have no nilpotent elements (\ref{602a},\ref{602b}) is certainly the correct
result for $d=2$, but it also shows that $K_i(\A_{\tl})$ and  $K_i(\A_{\tl'})$
do not contain
nilpotent elements and therefore (\ref{602a},\ref{602b}) holds in any
dimension.
The result becomes more and more complex  for higher $d$.
The same result could be obtained by iterative application of the
Pimsner-Voiculescu sequence onto the r.h.s.\ of (\ref{6pa1}), provided
this sequence splits at a certain position. This in general true for $d=2$
and has been used for the computation in \cite{Els1} and \cite{Kel};
for $d=3$ the splitting property has been explicitly checked in \cite{Els2}.
The latter approach is also useful in cases where $\Om$ is not of the
form of a Cartesian product.
\bs

Despite of the complicated structure of the $K$-group for higher dimensions
the image of a state on the $K_0$-group which is induced from a
 tracial product state on $\A_\tl$ is a simple expression. A tracial
 product state
is a trace that factorizes,
i.e.\ $\tr=\tr_1\otimes\tr_2\otimes\cdots\tr_d$ where $\tr_k$ is a trace on
$\A_{\tl_k}$.
In the non periodic case $\A_\tl$ is isomorphic to an iterated crossed product
by $\Z$:
\be           \label{6pa1}
\A_\tl = C(\Om)\times_\varphi\Integer^d = (C(\Om)\times_{\varphi_1}\Z)\cdots
\times_{\varphi_d}\Z .
\ee
By a little
abuse of notation the corresponding $\Z$-action on
$(C(\Om)\times_{\varphi_1}\Z)\cdots\times_{\varphi_{k-1}}\Z$
is also denoted by $\varphi_k$. This form for $\A_\tl$ is useful for the proof
of the following
theorem.
\bt
Let $\tl$ be a $d$-fold Cartesian product of one dimensional
non periodic tilings $\tl_k$
 and $\tr$ a tracial product state on $\A_\tl$. Then
\be                                    \label{603}
\tr_* K_0(\A_\tl)
 = \mu(C(\Omega,\Integer)) ,
\ee
$\mu=\mu_1\times\cdots\times\mu_d$ being the invariant measure determined by
the trace.
\et

{\em Proof:}
This theorem is an application of Theorem~3 of \cite{Pi} which we first cite:
Let $\tr$ be a trace on the crossed product
$A\times_{\alpha}\Integer$ of $\Z$ with
an arbitrary \CA. It restricts to a trace on $A$ which is also denoted by
$\tr$. Then the map
$\underline{\Delta}^{\alpha}_{\tr}:ker(id-\alpha_*)\subset K_1(A)
\longrightarrow\Real/\tr_*(K_0(A))$ defined by
\begin{equation}
\underline{\Delta}^{\alpha}_{\tr}([u]) =
\frac{1}{2\pi i}\int_0^1 \tr(\dot{\xi}(t)\xi^{-1}(t))\ dt\:/\:
\tr_*(K_0(A)) ,
\end{equation}
$\xi:[0,1]\longrightarrow GL(A)$ being a piecewise smooth path from
$1$ to $u\,\alpha(u^{-1})$, is a well defined group homomorphism, and
moreover  the sequence
\begin{equation}                                  \label{604}
0\longrightarrow \tr_*(K_0(A))\longrightarrow
\tr_*(K_0(A \times_{\alpha}\Integer))
\stackrel{q}{\longrightarrow}
\underline{\Delta}^{\alpha}_{\tr}(ker(id-\alpha_*))\longrightarrow 0 ,
    \nonumber
\end{equation}
$q:\Real\longrightarrow\Real/\tr_*(K_0(A))$ being the canonical
projection, is exact. Hereby the restriction of $\tr$ to $A$ is also denoted
by $\tr$.

We shall apply that theorem to $A=\A_{\tl'}\otimes C(\Omega_d)$ and
$\alpha=\varphi_d$ where
$\tl'=\tl_1\times\cdots\times\tl_{d-1}$, because
$\A_{\tl'}\otimes C(\Omega_d)\times_{\varphi_d}\Integer = \A_\tl$.
For this we need to determine
$\ker(id-\varphi_{d\,*})\subset K_1(A)$.
Representatives of elements of $K_1(A)$ are continuous functions
$u : \Omega_d \rightarrow GL(\A_{\tl'})$ satisfying
$\varphi_{d\,*}[u] =[\varphi_d^*u] = [u\circ\varphi_d^{-1}]$.
Hence
$(id-\varphi_{d\,*})[u] = 0$ whenever $u\sim u\circ\varphi_d^{-1}$.
By transitivity of $\varphi_d$ this implies
$\forall x,y\in\Omega_d : u(x)\sim_{\A_{\tl'}}u(y)$, the equivalence
here being homotopy equivalence in $GL(\A_{\tl'})$.
Since the fundamental group of totally disconnected spaces
is trivial,
the homotopies in $GL(\A_{\tl'})$ at different points
may be put together to yield a homotopy in  $GL(A)$ between $u$
and the constant function $\tilde{u}(x) = u(x_0)$ for some $x_0$, i.e.\
$\tilde{u}\sim u$. Hence any element of
 $\ker(id-\varphi_{d\,*})$ may be represented by a constant function over
$\Omega_d$ and $\ker(id-\varphi_{d\,*})\cong K_1(\A_{\tl'})$.
Now  $\tilde{u}\varphi_d^*(\tilde{u}^{-1}) = 1$ implies
$\underline{\Delta}^{\varphi_d}_{\tr}([u]) = 0$ so that with (\ref{604})
\begin{equation}                          \label{605}
\tr_*K_0(\A_{\tl'}\otimes C(\Omega_d)\mal{\varphi_d}{}) =
\tr_*K_0(\A_{\tl'}\otimes C(\Omega_d))
\end{equation}
Now let $[x]\in K_0^+(\A_{\tl'}\otimes C(\Omega_d))$.
As $[x]$ is represented by a projection valued continuous function $x$ over
$\Om_d$
there is a partition
$\{U_i\}_{i\in I}$ of $\Om_d$ by disjoint closed and open subsets (which has to
be finite) such that $x$ is equivalent to a projection of the form
$\sum_{i\in I}p_i\otimes\Ch_{U_i}$,
$p_i^2=p_i\in M_\infty(\A_{\tl'})$ and $\Ch_{U_i}$ being
the characteristic functions. Therefore
\be
\tr_*([x])=\sum_{i\in I}\int_{U_i} d\mu_d\,\tr (p_i) =
\sum_{i\in I}\mu_d(U_i)\, \tr_*([p_i])
\ee
where the trace on $M_\infty(\A_{\tl'})$ has also been denoted by $\tr$.
Hence
\be
\tr_*K_0(\A_{\tl'}\otimes C(\Omega_d)) =
\mu_d(C(\Omega_d,\Integer))\:\tr_*K_0(\A_{\tl'}) ,
\ee
and successively the theorem follows.\eb\bs

Again, if $\mu_k$ are ergodic measures, the tracial product state on $\A_\tl$
satisfies the Shubin formula.

\section{Substitution tilings}

In section~\ref{sec3} the results obtained for $K_0(\A_\tl)$
resp.\ $\tr_*(K_0(\A_\tl))$ have still been rather abstract.
To compute explicitly these groups
or an invariant measure on the hull of $\tl$ extra structure is needed. Such a
structure is given for certain non periodic self-similar tilings by a \sst,
the corresponding tilings called \sst\ tilings (or similarity
tilings).
Substitutions are not only useful for Cartesian products of one dimensional
tilings but also for
tilings with nontrivial orientational symmetry $\Sy$.
In these cases, in which $K_0(\A_\tl)$ is unknown to us, the \sst s
are used to determine the part of $\tr_*(K_0(\A_\tl))$ which is given by
an invariant
measure on the hull. For the \sst\ to be
well defined none of its \ti s should be invariant under $\Sy$.
This might require additional markings for breaking the symmetries.\bs
\newcommand{\sa}[1]{F(a_{#1})}

A {\em \dfl} $\rho_\tat$
of a tiling $\tl$ is a {\em local} prescription by which each \ti\ of  $\tl$
is replaced by a pattern,
made from representatives of the original \tiii s rescaled by $\tat^{-1}$,
in such a way that these patterns fit together to
form a new tiling $\rho_\tat(\tl)$
which differs from $\tat^{-1}\tl$ only by a translation.
The tiling $\tat^{-1}\tl$ is $\tl$ rescaled by $\tat^{-1}$, $\tat>1$,
where $0\in\Real^d$ is the point to be kept fixed,
so it has the same \tiii s as $\rho_\tat(\tl)$, and {\em local}
refers to the following requirement which a \dfl\ has to satisfy. \bi
\item[\bf 1)]
If the tiles at $x\in\tl\2$ and at $x'\in\tl\2$
differ only by a translation and the one at $x$ is
replaced by the pattern $(M,y)\subset (\rho_\tat(\tl),z)$ then the one
at $x'$ is to be replaced by $(M,y)\subset (\rho_\tat(\tl),z-(x\!-\!x'))$.
\ei
The \mi\ by which \ti\ $a$ of $\tl$ is replaced is denoted by
$\rho_\tat(a)$.
Furthermore it will be required that $\rho_\tat$ is $\Sy$-covariant,
i.e.\ that for a \ti\ $a$
\bi
\item[\bf 2)]
$\rho_\tat(g\cdot a)$ and $g\cdot\rho_\tat(a)$ are in the same \mii.
\ei
A \dfl\ of $\tl$ is then determined by the \miii es of the $\rho_\tat(a)$
together with finitely many positions, namely at most one for one
representative of each \tii. In particular
 $\rho_\tat$ may be understood as a map from the set of \miii es of $\tl$
into the set of rescaled \miii es. In fact the position of $\rho_\tat(\tl)$
in $\Real^d$ is irrelevant for the sequel.
Note that we do not require the
 \mi\  $\rho_\tat(a)$
to fill out exactly the space of the original \ti\ $a$
-- if it does it is a decomposition in the sense of \cite{GrSh} --
but {1)} implies that $\lim_{r\rightarrow\infty}
\frac{\mbox{\scriptsize vol}\,\rho_\tat(M_r(\tl))}
{\mbox{\scriptsize vol}\,M_r(\tl)}=1$, $\mbox{vol}M$ denoting the volume of the
subset of $\Real^d$ covered by the \mi\ $M$.
Thus the volume of $\rho_\tat(a)$ is in a certain
sense of average equal to the volume of $a$.
This fixes the scaling factor
$t$.

It is not difficult to extend the action of $\rho_\tat$ to all elements
of ${\cal L}'_\tl = \overline{\{\tl-x|x\in\Real^d\}}$.\footnote{
If $\tl$ is homogeneous then ${\cal L}'_\tl = {\cal L}_\tl$.}
 Setting $\rho_\tat(\tl-x)=\rho_\tat(\tl)-x$ one obtains
a surjective map $\rho_\tat:\{\tl-x|x\in\Real^d\}\rightarrow
\{\tat^{-1}\tl-x|x\in\Real^d\}$ which is continuous by {1)} and may
therefore
be extended to the closures
%${\cal L}'_\tl = \overline{\{\tl-x|x\in\Real^d\}}$ resp.\
${\cal L}'_\tl$ resp.\
${\cal L}'_{\tat^{-1}\tl}$.
Clearly, not every tiling admits a \dfl\ but its existence
is a special property which
in particular expresses a kind of self-similarity (for this reason
\sst\ tilings are called similarity tilings in \cite{GrSh}).

A {\em \sst} $\rho$ of a tiling $\tl$
is a \dfl\ followed by a rescaling by the
factor of $\tat$ so that the \ti s of $\rho(\tl)$ have original size.
Hence, if a tiling $\tl$ is a \sst\ tiling, i.e.\ if it
allows for a \sst\ $\rho$ -- it will also be called
invariant under the \sst\ -- then $\exists x\in\Real^d: \rho(\tl)=\tl-x$.
We call a pattern $\rho(a)$ of $\rho(\tl)$
which is the replacement of the tile $a$ of $\tl$
a {\em \fl}.

The right inverse of $\rho_\tat$ as a map
$\rho^{-1}_\tat:{\cal L}'_{\tl}\rightarrow {\cal L}'_{\tat\tl}$,
$\rho_\tat\circ \rho^{-1}_\tat=\mbox{id}$, is called
an {\em \ifl}\footnote{
Here we follow the terminology established in the physical literature which
unfortunately
differs from the one in \cite{GrSh} where the term \ifl\ is used for what
is called \sst\ here.}
and
correspondingly, the inverse of a \sst\ is an \ifl\ followed by a rescaling
by $\tat^{-1}$. However a specific \dfl\ may have several right inverses,
if it is not injective. Uniqueness of the right inverse enforces non
periodicity as is
shown in \cite{GrSh}:
\bl
If a \dfl\ $\rho_\tat$ of $\tl$ has a unique right inverse then $\tl$ is non
periodic.
\el
{\em Proof:} Assume that $\tl-x=\tl$ for a $x\in \Real^d$ such that $|x|$ is
minimal. If $\rho^{-1}_\tat$ is a right inverse of $\rho_\tat$ then
$\rho^{-1}_\tat$ followed by a translation by $x$ is as well a right inverse.
Uniqueness then implies that $\rho^{-1}_\tat(\tl)$ is invariant under
translation by $x$ and therefore $\tl-\tat^{-1}x=\tl$ which contradicts
the minimality of $|x|$.\eb\bs

It is not at all clear that an \ifl\  can be formulated as a local procedure,
since it has to be determined whether a given \mi\ of
$\rho(\tl)$ is a \fl\ of some \ti\ $a$ of $\tl$;
this is not insured by the \mi\ being congruent to $\rho(a)$,
since e.g.\ all its neighbouring
\ti s have to form \fl s, too.
We call an \ifl\ {\em local} and correspondingly a \sst\ {\em locally
invertible}, if the determination of whether or not a \mi\ of
$\rho(\tl)$
is a \fl\ of a \ti\ of $\tl$ may be uniquely carried out by
inspection of a patch of given finite size containing that \mi.
More technically a \sst\ is locally invertible, if there is an $r$ such that
for all $x\in\rho(\tl)\2$ the \fl\ of which $x$ is a \pkt\ is uniquely
determined by the pattern $M_r(\rho(\tl)-x)$.
It will be henceforth required that
\bi
\item[\bf 3)]
$\rho$ is locally invertible.
\ei
In particular a locally invertible \sst\ has a unique right
inverse, so that we are dealing with non periodic tilings.
The important point of local invertibility of a \sst\ is
that we can for any $T\in\Om$ uniquely determine the \fl s by inspection of
finite patches and moreover repeat this process and determine the $n$-fold
\fl s,
i.e.\ the \mi s which correspond to $n$-fold replacements $\rho^n(a)$ of
\ti s $a\in\rho^{-n}(T)$. Of course, the size of the \mi\ needed to determine
the $n$-fold \fl\ to which a \ti\ belongs has to grow exponentially with $n$.
This may be used to construct a continueous
map $\jo$ from the hull of $\tl$ onto the path space of a graph
$\Sigma$ being related
to the \sst\ as follows.\ms

The \sst\ matrix $\sigma$ of the \sst\ $\rho$ has entries
\be
\sigma_{ij} =
\hbox{number of representatives of $[a_j]$ in $\rho(a_i)$}
\ee
$[a]$ denoting the equivalence class under all Euclidean transformations of the
tile $a$.
These entries are all  positive and integer and therefore $\sigma$ may be
interpreted as
the \cm\ of a graph $\Sigma$. The vertices $\Sigma^{(0)}$ of this graph are in
one to one correspondence with the \tiii s  and
vertex $i$, which corresponds to $[a_i]$, is linked to vertex $j$ by
$\sigma_{ji}$ oriented
edges.  Hence an (oriented) edge $\k$ has a source
$s(\k)$ and a range $r(\k)$. Two edges $\k_1$,
$\k_2$  may be concatenated, denoted by
$\k_1\circ\k_2$ or for short by $\k_1\k_2$, if
they fit together, i.e.\  if $r(\k_1)=s(\k_2)$. A
(finite or infinite) path over $\Sigma$ is a (finite or
infinite) sequence of concatenated edges which fit together. One may
then analogously define the source and (if the path is finite)
the range of a path as the source of the first and the range of
the last edge as well as concatenation of two paths.  We denote
by $\pf^{(n)}$ the space of all paths of length $n$ and by $\pf$
all infinite paths over $\Sigma$.
$\pf$ is also called the path space over
$\Sigma$. It
becomes a metric space if one defines the distance of two paths
to be $d(\gamma,\gamma')=\exp(-l(\gamma,\gamma'))$ with
$l(\gamma,\gamma')=\sup\{i\geq 0|\forall j\leq
i:\gamma_j=\gamma'_j\}$
where $\gamma_i$ denotes the $i$'th edge of $\gamma$.
The metric topology is generated by
sets which are labelled by a finite path $\xi$ \be
U_\xi=\{\xi\circ\gamma|\gamma\in\pf,s(\gamma)=r(\xi)\} . \ee
These sets are open and closed which implies that $\pf$ is
totally  disconnected. An argument similar to the one above
for the hull shows that $\pf$ is compact. \ms

The \sst\ is called {\em primitive} if $\sigma$ is primitive (or
aperiodic), i.e.\ if some power of $\sigma$ has only strictly
positive entries. In this case $\Sigma$ is connected and by the
Perron-Frobenius theorem  $\sigma$ has a non degenerate largest
eigenvalue (the \pfw ) whose corresponding left and right
eigenvectors (\pfv s) may be chosen to have only strictly
positive entries.
The $i$'th component of the left \pfv\ $\nu$, normalized to
$\sum_j\nu_j=1$, furnishes the relative frequency
of representatives of \tiii\ $[a_i]$ in $\tl$.
Therefore the \pfw\ is $\tau=\tat^d$,
the scaling factor to the power of the
dimension of the tiling by which the \sst\ differs from the \dfl,
i.e.\ the volume of the \mi\ $\rho(a)$ is in average $\tau$ times larger than
that of $a$.
We consider only primitive \sst s.

\subsection{The map $\joh$ onto the path space}

The \sst\ $\rho$ can be used to  construct a continuous surjective mapping
from
 the hull $\Om$ onto the  path space $\pf$.  The
$\sigma_{ji}$ edges linking vertex $i$ and $j$ may be assigned
to the $\sigma_{ji}$ different positions a representative $a_i$
of $[a_i]$ may have in $\rho(a_j)$; and
for this to be well defined it is crucial that no \ti\ is invariant under
$\Sy$.
Thus such an edge uniquely
encodes that position  and by interpreting the $l$'th edge $\k_l$ of a path
$\k_1\k_2\cdots\k_n$ as encoding the
position of a representative of $[a_{s(\k_l)}]$ in $\rho(a_{r(\k_l)})$
the whole path
 encodes the position of a representative of
$[a_{s(\k_1)}]$ in $\rho^n(a_{r(\k_n)})$.
For a Penrose tiling this is illustrated in figure~\ref{b2}.
Now
let $T\in\Om$, $a_{i_0}$ be the \ti\ on $0$,  and choose an
increasing chain $a_{i_0}\in\eta_1\subset\eta_2\subset
\cdots$ of \mi s which approximate $T$ such that $\eta_n$ is
sufficiently large
to determine the $n$-fold \fl\ in $T$ which covers $0$.
This is possible by the local invertibility of $\rho$
 the size of $\eta_n$ having to grow exponentially.
The $n$-fold \fl, to which in particular $a_{i_0}$ belongs, is
the $n$-fold \fl\ of some \ti\ $a_{i_n}$.
Let $\jo_n(T)$ be the path
%of length $n$ starting at $i_0$ and ending at $i_n$
which encodes
the position of $a_{i_0}$ in that $n$-fold \fl\ $\rho^n(a_{i_n})$.
It is clear that the
first $n\!-\!1$ edges of $\jo_n(T)$ yield $\jo_{n-1}(T)$ so that
we may define $\jo(T)$ to be the path whose
$n$'th edge is given by the $n$'th edge of $\jo_n(T)$.

Decompose $\jo(T)=\jo_n(T)\circ\gamma$.
Denote by $\Mi{n}{T}$
the \mii\ of $\rho^n(a_{i_n})$ with $a_{i_n}$ as above
 and by
$\pk{n}{T}\in \Mi{n}{T}\2$
the \pkt\ of the \ti\ that is encoded by $\jo_n(T)$.
Thus $\Mi{n}{T}$ depends only on $r(\jo_n(T))=i_n$ and the orientation of
$a_{i_n}$ which is determined by $T$.
Furthermore let $\xi\in\pf^{(n)}$ be any other path with $r(\xi)=i_n$
and let $x_\xi\in\Mi{n}{T}\2$ denote the \pkt\ of the \ti\ in
$\Mi{n}{T}$ which is encoded by $\xi$.
Then $(\Mi{n}{T},\pk{n}{T})\subset (T,0)$ and
$(\Mi{n}{T},x_{\xi})\subset (T-(x_{\xi}\!-\!\pk{n}{T}),0)$, so that
\be  \label{5pa1}
\jo(T-(x_\xi\!-\!\pk{n}{T}))=\xi\circ\gamma.
\ee
\bt
$\jo:\Om\rightarrow\pf$ is a continuous surjective map.
\et
{\em Proof:}
$\jo$ is continuous if
$\forall\epsilon>0\,\exists\delta>0:\,d(T,T')<\delta \Rightarrow
d(\jo(T),\jo(T'))<\epsilon$, which here amounts to
$\forall L>0\,\exists R>0:\,r(T,T')>R \Rightarrow
l(\jo(T),\jo(T'))>L$.
But this follows from the local invertibility of $\rho$, since
$l(\jo(T),\jo(T'))>L$ if and only if
$\jo_{L+1}(T)=\jo_{L+1}(T')$.
Surjectivity of $\jo$ follows from compactness of $\Om$:
Given $\gamma\in\pf$ let $\{\gamma^{(n)}\}_{n\geq 0}$ be a
sequence of infinite paths such that
the first $n$ edges of   $\gamma^{(n)}$ coincide with the first
$n$ edges of  $\gamma$ but
the $m$-th edges for  $m>n+m_0$ with the one of $\jo(\tl)$ --
by the primitivity of the \sst\ this is possible for some finite
$m_0$. By (\ref{5pa1}) there is a sequence of tilings $\{T^{(n)}\}_{n\geq 0}$
which are translates of $\tl$ such that
$\jo(T^{(n)})=\gamma^{(n)}$. Since  $\Omega$ is compact
$\{T^{(n)}\}_{n\geq 0}$ has a
convergent subsequence and its limit is a preimage of
$\gamma$.\eb
\bs

Certainly, $\jo$ cannot be injective if the tiling has a
nontrivial orientational symmetry, because
of $\jo(g\cdot T)=\jo(T)$. But if we pass over to the quotient
space (with quotient topology) and define
$\joh:\Om/\Sy\rightarrow\pf$ by $\joh([T])=\jo(T)$
a criterion for the invertibility of $\joh$ may be given.
For this to become clear the notion of a singular tiling is useful.

The inner radius of a \mi\ $(M,x)$ is the radius of the largest ball around
$x$ that is covered by $(M,x)$.
Hence, if $x$ is close to the boundary of $M$, the inner radius will be
relatively small no matter how large $M$ is.
A tiling $T\in\Om$ is called {\em regular}, or more precisely
{\em $\rho$-regular},
if the inner radius of $(\Mi{n}{T},\pk{n}{T})$ diverges with $n$,
otherwise it is called {\em singular}.
As $\Mi{n+1}{T}$ is congruent to $\rho^n(\rho(a_{i_{n+1}}))$ it contains
$\Mi{n}{T}$ leading to the increasing chain of
finite patterns
\be                       \label{x1}
(\Mi{n}{T},\pk{n}{T})\subset
(\Mi{n+1}{T},\pk{n+1}{T})\subset\cdots\subset (T,0)
\ee
which is called the {\em \ap} of $T$. Up to the orientation the \ap\ of $T$ is
completely determined by $\jo(T)$.
If $T$ is regular its
\ap\ is an approximation of it in the sense that any \ti\ of it
lies in some $(\Mi{n}{T},\pk{n}{T})$.
In contrast singular tilings may or may not
be determined by their \ap s,
as one needs to know, what
happends outside the inner radius of the limit of the \ap s.
Singular tilings do in fact exist:
\bl
If $T$ is symmetric then it is singular.
\el
{\em Proof:}
Assume $\exists x\in T\2:\,g\cdot T = T-x$, hence $\jo(T)=\jo(T-x)$.
%As the \tiii s have no symmetry  $x\neq 0$.
Let us for a moment identify $\Mi{n}{T}\2$ via (\ref{x1})
with a subset of $T\2$.
Assume
$x\in \Mi{n}{T}\2$. Then $x$ is the \pkt\ encoded by
$\jo_n(T-x)$ in $\Mi{n}{T}$.
But $\jo_n(T-x)=\jo_n(T)$ implies that $x=\pk{n}{T-x}=\pk{n}{T}=0$
contradicting
the requirement that the \tii s do not have symmetries.
Hence $x\notin \Mi{n}{T}\2$ implying that the
inner radius of $(\Mi{n}{T},\pk{n}{T})$ is smaller than
$|x|$.\eb\bs

Given a \mi\ $M$ of some tiling $T$ we define the {\em \saum}
of $M$ in $T$ to be the pattern of
\ti s in $T$ which touch $M$
but do not belong to it.
If $T$ contains some underlying structure one expects
an \mii\ of $T$ to force an occasionally larger \mii\ $\tilde{M}$
 in the sense that whenever $(M,x)\subset (T,y)$ then also
$(\tilde{M},x)\subset (T,y)$.
We say that $\rho$ {\em forces its \saum} if there is an $N$ such
that whenever
the $N$-fold \fl\ $\rho^N(a_i)$ occurs in $T$
its \saum\ is always the same; we
denote the \mii\ of $\rho^N(a_i)$ together with the \saum\ occuring in
the above case by $\sa{i}$.
This requirement is weaker than the condition that the \mii es of
the $\rho^N(a_i)$ all force their \saum s, since we require its \saum\ only
to be the same if the \mi\ is an actual
 $N$-fold \fl.
Now if the $N+1$-fold \fl\ $\rho^{N+1}(a_i)$ occurs in $T$ then even
$\rho(\sa{i})$
has to occur at that place.
This simply follows from the fact that
 in that case $\rho^N(a_i)$ occurs in
$\rho^{-1}(T)$ as a $N$-fold \fl\
and forces $\sa{i}$ to occur there.
%the surrounding \ti s which have at least one vertex with $\rho^N(a_i)$ in
%common form always the same pattern.
It will be shown that if
\bi
\item[\bf\bii] $\rho$ forces
its \saum
\ei
$\joh$ is indeed a homeomorphism.
\bl                   \label{lemb1}
If $\rho$ forces its \saum\ then all $T\in\Om$ are uniquely determined
by their \ap.
\el
{\em Proof:}
Let $T\in\Om$ perhaps be singular and consider its \ap. As above set
$i_n=r(\jo_n(T))$.
The \mi\ $(\Mi{N}{T},\pk{N}{T})\subset (T,0)$ is by construction the
$N$-fold \fl\ which covers $0$ so that
by \bii\ for all $m\geq 0$
\be
(\Mi{N+m}{T},\pk{N+m}{T})\subset (\rho^m(\sa{i_{N+m}}),\pk{N+m}{T})\subset
(T,0)
\ee
where the orientation of $a_{i_{N+m}}$ is determined by $\Mi{N+m}{T}$.
Therefore the \ap\ determines in fact the increasing chain
\be          \label{18042}
%(\sa{i_{N}},\pk{N}{T})\subset
(\rho^m(\sa{i_{N+m}}),\pk{N+m}{T})\subset
(\rho^{m+1}(\sa{i_{N+m+1}}),\pk{N+m+1}{T})\subset\cdots\subset (T,0) .
\ee
Let $y\in \sa{i_N}\2$ and
take the translate
$T_y=T-(y\!-\!\pk{N}{T})$, i.e.\ $(\sa{i_N},y)\subset (T_y,0)$.
Then $(\Mi{N+m}{T_y},\pk{N+m}{T_y})\subset
(\rho^m(\sa{i_{N+m}}),\pk{N+m}{T}-y)$
showing that the \ap\ of $T_y$ is already determined by (\ref{18042}) and hence
by the  \ap\ of $T$.
Since $(\sa{i_N},\pk{N}{T})\subset (T,0)$ contains at least all
\ti s touching $a_{i_0}$,
the \ti\  on $0$, we may show by induction,
namely repeating the same argument with $T_y$ in place of $T$, that the
\ap\ of $T$ already determines the \ap s of all its translates, and hence all
of
$T$.\eb

\bt   \label{t4}
If $\rho$ forces its \saum\ then $\joh:\Om/\Sy\rightarrow\pf$ is a
homeomorphism.
\et
{\em Proof:}
Continuity and surjectivity of $\joh$ follow from the continuity
and surjectivity of $\jo$, and since
$\Om/\Sy$ as well as $\pf$ are compact we have to show that
$\joh$ is invertible.

Let $\jo(T)=\jo(T')$.
In particular $\jo_n(T)=\jo_n(T')$ which by covariance of $\rho$
implies that there is a
$g\in\Sy$ such that
$g\cdot a_{k_0}=a'_{k_0}$ and $g\cdot
\rho^n(a_{k_n})=\rho^n(a'_{k_n})$. Hence
$(g\cdot \Mi{n}{T},g\cdot \pk{n}{T})=(\Mi{n}{T'},\pk{n}{T'})$
implying by Lemma~\ref{lemb1}
$g\cdot T=T'$.\eb
\bs

The last two theorems show that for a tiling which is invariant under a locally
invertible \sst\ that forces its \saum\
 not only $\A_\tl$ is naturally assigned to them but also the $AF$-algebra
$\AF$ of the path space; c.f.\ the appendix for its definition.
One advantage of this formulation is that it allows us to compute
the image of an $\Gr$-invariant $\Sy$-invariant measure on $\Om$.
But before coming to that a sufficient criterion for $\tl$
to be homogeneous will be given.

We call a \sst\ $\Sy$-primitive if there is an $N$ such that for any \tii\ $a$
any other \tii\ $b$ occurs in $\rho^N(a)$.
This implies the existence of
an $r$ such that for any $x\in\tl\2$ any \tii\ occurs
in $M_r(\tl\!-\!x)$. Hence if $\tl$ allows for a $\Sy$-primitive \sst\
then any \tii\ occurs in any $T\in\Om_\tl$ and consequently also
any $n$-fold \fl\ in any orientation.

\bl \label{l5}
Let $\tl$ be invariant under a $\Sy$-primitive \sst\ which satisfies \bii.
Then $\tl$ is homogeneous.
\el
{\em Proof:}
Let $T\in \Om_\tl$ and $M$ be an \mii\ of $\tl$, we have to show that
$\exists y'\in T\2:(M,x)\subset (T,y')$. Under the hypothesis $\joh$ is a
homeomorphism and therefore exists an $n$ and $I\subset \pf^{(n)}$ such that
$\jo(U_{M,x})=\bigcup_{\xi\in I} U_\xi$.
Finiteness of the union of the r.h.s.\ follows from compactness
of $\jo(U_{M,x})$ and disjointness can then always be achieved so that one can
find such an $n$.
The statement $\exists g\in\Sy:(g\cdot M, g\cdot x)\in (T,0)$ is thus
equivalent to
$\jo_n(T)\in I$. Primitivity of the \sst\ now implies that for all $\xi\in I$
there is a $y\in T\2$ such that $\jo_n(T-y)=\xi$, and hence
$\exists g\in\Sy:(g\cdot M, g\cdot x)\in (T,y)$. But we need $\Sy$-primitivity
to insure that there is as well a $y'\in T\2$ such that
$M_n(T-y')=g^{-1}\cdot
M_n(T-y)$ and hence $(M,x)\in (T,y')$.
\eb

\subsection{The $\Sy$-invariant measure on $\Om$}

Since the operators of the form (\ref{22pa1}) depend only on the \miii es and
not on their orientation
they are invariant under $\Sy$, i.e.\ they are elements of
 $\A_\tl^\Sy:=\{f\in\A_\tl|f(g\cdot T,g\cdot S) = f(T,S)\}$.
A trace on $\A_\tl^\Sy$ may be seen as a $\Sy$-invariant trace on $\A_\tl$ and
restricts  to a $\Gr$-invariant $\Sy$-invariant measure on $\Om$ and it is
this measure which may be determined using the structure of the path space
$\pf$.\ms

Remember that by
the lack of symmetry of the \tiii s
\be                            \label{332}
U_{M,x}\cap U_{g\cdot M,g\cdot x}=\emptyset .
\ee
For this reason a one to one correspondence of
measures $\tilde{\mu}$ on $\Omega/\Sy$ with $\Sy$-invariant measures
 $\mu$ on $\Omega$ is given on the images of sets $U_{M,x}$ under the natural
projection $[\cdot]:\Om\rightarrow\Omega/\Sy$ by
\be                                     \label{5pa2}
\tilde{\mu}([U_{M,x}]) = \mu(\bigcup_{g\in\Sy}U_{g\cdot M,g\cdot x}) =
|\Sy|\,\mu(U_{M,x}).
\ee
The factor $|\Sy|$ appears, as the topology on $\Om$ was defined through
oriented \mi s and not through their $\Sy$-orbits.
In fact, the restriction to $\Sy$-invariant operators is not crucial here
and we could in principle allow for any operator being represented by an
element
of $\A_\tl$ but
one might expect that traces of $\Sy$-invariant projections take their values
in $|\Sy|\mu(C(\Om,\Z))$.

$\AF$ is actually a groupoid-\CA, too.
The groupoid $\Gr_\Sigma$ is defined by the
equivalence relation $\gamma\sim_{\Gr_\Sigma}\gamma'$ whenever $\exists
n_0\forall n\geq n_0:
\gamma_n=\gamma'_n$ and its topology is
generated by $U_{\xi,\xi'}:=\{(\xi\circ\gamma,\xi'\circ\gamma)|
\gamma\in\pf,s(\gamma)=r(\xi)\}$ where $\xi$ and
$\xi'$ have the same finite length and $r(\xi')=r(\xi)$. It leads as well
to discrete
orbits $\Gr_\Sigma^\gamma$.

The primitivity of $\sigma$ implies that $\AF$ has a unique (normalized) trace
and hence there is a unique (normalized) $\Gr_\Sigma$-invariant measure
$\mu_\Sigma$ on $\pf$. $\Gr_\Sigma$-invariance in this case means that
$\mu_\Sigma(U_\xi)$ depends
only on $r(\xi)$, c.f.\ (\ref{4pa1a},\ref{4pa1b}).
\bt              \label{thmb1}
Let $\mu$ be a (normalized)
$\Gr$-invariant $\Sy$-invariant measure on $\Om$, the hull of
a \sst\ tiling and $\tr$ the unique (normalized) trace on the corresponding
$AF$-algebra $\AF$.
Then
\be                                                 \label{3312}
\tr_*K_0(\AF) \subset
\tilde{\mu}(C(\Omega/\Sy,\Integer))
\ee
$\tilde{\mu}$ being given by (\ref{5pa2}),
and if the \sst\ satisfies \bii\ both sets are even equal.
\et
{\em Proof:} $\tilde{\mu}$ induces a measure $\hat{\mu}$ on $\pf$ by
\be
\hat{\mu}(U_\xi) = \tilde{\mu}(\joh^{-1}(U_\xi))
=\mu(\{T\in\Om|\jo_n(T)=\xi \}),
\ee
$n=|\xi|$.
Whereas $T\sim_\Gr T'$ does not
imply $\jo(T)\sim_{\Gr_\Sigma}\jo(T')$, the converse
 is true by (\ref{5pa1}).
Fix one $T'\in\Om$ such that $\jo_n(T')=\xi$ and let $\xi'$ be a path of length
$n$ with $r(\xi')=r(\xi)$ then again using
 (\ref{5pa1})
\begin{eqnarray}
\mu(\{T|\jo_n(T)=\xi \})&=&|\Sy|\mu(\{T|
(\Mi{n}{T'},\pk{n}{T'})\subset(T,0)\:
\wedge\:\jo_n(T)=\xi \}) \nonumber \\
&=&|\Sy|\mu(\{T+(x_{\xi'}\!-\!\pk{n}{T'})|(\Mi{n}{T'},x_{\xi'})\subset(T,0)\:
\wedge\:\jo_n(T)=\xi' \}) \nonumber \\
&=&\mu(\{T|\jo_n(T)=\xi' \}),\nonumber
\end{eqnarray}
the last equality following from the $\Gr$-invariance of $\mu$,
and hence $\hat{\mu}$ is $\Gr_\Sigma$-invariant.
But since the trace on $\AF$ is unique up to normalization
$\hat{\mu}=\mu_\Sigma$.
As for $AF$-algebras $\tr_*K_0(\AF) =\mu_\Sigma(C(\pf,\Integer))$ the inclusion
is proven. Clearly
$\mu_\Sigma(C(\pf,\Integer))=\tilde{\mu}(C(\Om/\Sy,\Integer))$ if
 $\joh$ is a homeomorphism.\eb\bs

As an application we compute the invariant measure on the hulls of certain
generalizations of the Fibonacci chain and of a Penrose tiling.

\subsubsection{Examples: Generalizations of the Fibonacci chain and Penrose
tilings} \label{s421}

We first give examples for one dimensional \sst s which have been extensively
studied over the past. Their invariant measure on the hull may as well be
computed by methods developed in \cite{Q}. But we prefer to use the $K_0$-group
of $\AF$ since it not only provides us with the values of the invariant measure
but also yields a finer labelling in case $\tr_*$ is not injective.
This is based on the equality of $K_0(\A_\tl)$ and $K_0(\AF)$ in the one
dimensional case the proof of which is postponed to subsection~\ref{sec5}.
As a two dimensional example we determine the invariant measure on the hull of
a Penrose tiling.\bs

A rather efficient way of thinking of an $n$-letter \sst\ is as a homomorphism
$\rho$ of the free group of $n$ generators $F_n$ \cite{BGJ}. The letters
represent the generators and a tiling is a map $T:\Z\rightarrow F_n$ such that
formally the two-sided infinite
product $\prod_{i\in\Z} T_i$ is up to a finite shift of the indices invariant
under $\rho$.
Denote by $Ad_w$ the conjugation by word $w$, i.e.\ $Ad_w(a)=waw^{-1}$.
If $\rho(a)$ ends resp.\
begins for all $a$ on word $w$ then $Ad_w$ resp.\ $Ad_{w^{-1}}$ act like a
left resp.\ right shift on $\rho(a)$ and consequently
 $Ad_w\circ\rho$ resp.\  $Ad_{w^{-1}}\circ\rho$
lead to the same invariant sequences and the same hull as $\rho$.
This shall be used below, because the
\sst s considered satisfy \bii\ only after being iterated a few times
and then composed with an inner automorphism as above.
In fact we shall obtain \sst s of the form
$\tilde{\rho}=Ad_w\circ\rho^n$ which have the property that
\bi
\item[\bf\bii']
the first letter as well as
the last letter of $\tilde{\rho}(a)$ are independent of $a$.
\ei
This is sufficient for $\tilde{\rho}$ to satisfy \bii.

Let us mention here that
local invertibility of a \sst\ is insured by recognizability,
a concept introduced in \cite{Q}.
Local invertibility of $\rho$ implies local invertibility of $\tilde{\rho}$.
Moreover, since the \sst\ matrix of $\tilde{\rho}$ equals the $n$'th power of
the \sst\ matrix $\sigma$ of $\rho$,
the $AF$-algebra  defined by it has the same scaled ordered $K_0$-group as
$\A_\Sigma$ -- both algebras are in fact isomorphic --
and its determination may be carried out
just using the $\sigma$.
Here, the order unit is always $(1,\cdots,1)$ so that all left-\pfv s have to
be normalized to satisfy
$\sum_i\nu_i=1$.
Different letters are supposed to represent different congruence classes of
punctured intervals so that the orientational symmetry is trivial.
\subsubsection*{\A-type generalizations of the Fibonacci chain}

This family of \sst s is defined on the alphabet $\B=\{a_1,\cdots,a_k\}$,
$k\geq 2$, by
\be
\rho :\left\{\begin{array}{ll}
a_i\mapsto a_{k-i+1} a_{k-i} & \hbox{if }\:i\leq k-i  \\
a_i\mapsto  a_{k-i}a_{k-i+1} & \hbox{if }\:k-i<i<k  \\
a_k\mapsto a_1 &
\end{array}\right. .
\ee
and $k=2$ yields the Fibonacci chain.
An inverse procedure of $\rho$ may be locally defined as follows:
Starting with $i=1$ successively
 replace
any word of $T$ which is congruent to $\rho(a_i)$ by $\tilde{a}_i$.
This is a well defined procedure, since after the $i$'th such replacement the
result will not contain any more letter $a_{k-i}$. After the $k$'th such
replacement one ends up with a sequence
$\{\tilde{a_i}\}_{i\in\Z}$ which yields after removing the tildes a preimage
of $T$ under $\rho$.

$\rho$ does in this form not obey \bii. However if one considers only the
sequence of letters at the right end of $\rho^n(a_k)$ for $n\geq 0$ one obtains
$a_k\rightarrow a_1\rightarrow a_{k-1}
\rightarrow a_2\rightarrow a_{k-2}\cdots\rightarrow a_{[\frac{k+1}{2}]}
\rightarrow a_{[\frac{k+1}{2}]}\cdots$. Hence
$\rho^k(a_i)$ ends for all $i$ on $\rho(a_{[\frac{k+1}{2}]})=
a_{[\frac{k-1}{2}]}a_{[\frac{k+1}{2}]}$.
Therefore the
\sst\
 $\tilde{\rho}=Ad_{a_{[\frac{k+1}{2}]}}\rho^k$ satisfies \bii'.
The \sst\ matrix of $\tilde{\rho}$ is $\sigma^k$ where
\be
\sigma =\left( \begin{array}{ccccc}
0 & \cdots & 0 & 1 & 1 \\
0 & \cdots & 1 & 1 & 0 \\
  & \cdot & \cdot  &   &   \\
1 & 1 & 0 &\cdots & 0 \\
1 & 0 & \cdots & & 0 \end{array} \right) .
\ee
The corresponding graph is a tadpole graph. A twofold covering of it is
the Coxeter graph of the Coxeter group $\A_{2k}$, hence the name of the \sst.
It is invertible over $\Z$ so that
\be
K_0(\AF) = \Integer^k .
\ee
The \pfw\ of $\sigma$ is $\tau = 2\cos\frac{\pi}{2k+1}$
and the components of its
normalized left-\pfv\ are
$\nu_l = (2-\tau)\frac{\sin\frac{l\pi}{2k+1}}{\sin\frac{\pi}{2k+1}}$.
Since $2-\sigma$ is invertible over $\Z$,
\be
\tr_*K_0(\AF) =  \{\sum_{l=1}^k n_l\frac{\sin\frac{l\pi}{2k+1}}
{\sin\frac{\pi}{2k+1}}|n_l\in\Integer\}.
\ee
Moreover, $\tr_*$ is injective if and only if $2k+1$ is prime.
\subsubsection*{Metallic type generalizations of the Fibonacci chain}

The so called metallic means furnish the \pfw s of the \sst\ matrices
of the following class of 2-letter \sst s given by
\be
\rho :\left\{\begin{array}{l}
a\mapsto b^k a^l\\
b\mapsto a
\end{array}\right.
\ee
and parametrized by integers $k\geq 1, l\geq 1$.
They may be considered as alternative generalizations of the case $k=1$, $l=1$
which leads to the Fibonacci chains.
They are locally invertible: just replace any $b^k a^l$ by $a$ and the
remaining
$a$'s by $b$'s. But again they do not satisfy \bii\ in this form.
However if we
iterate them twice we obtain $a\mapsto  a^k(b^k a^l)^l$, $b\mapsto  b^k a^l$
so that
$\tilde{\rho}=Ad_{(b^k a^l)}\circ\rho^2$ satisfies \bii'.
The \sst\ matrix of $\tilde{\rho}$ is $\sigma^2$ where
\be
\sigma =\left( \begin{array}{cc}
l & k \\
1 & 0 \end{array} \right)
\ee
$\sigma$ has
 \pfw\ $\tau = \frac{l+\sqrt{l^2+4k}}{2}$ and normalized left-\pfv\
$\nu = \frac{1}{\tau+k}(\tau,k)=\frac{1}{k+l-1}(\tau-1,k+l-\tau)$.
Hence
\be \label{5bpa1}
\tr_*K_0(\AF)  =  \{\tau^{1-n}(n_1 \frac{\tau-1}{k+l-1}+n_2)|
n>0,(n_1,n_2)\in\Z^2\} ,
\ee
and since $\tau$ is irrational $\tr_*$ is injective so that $K_0(\AF)$
may be identified with its image under $\tr_*$.
It should be noted that $k\geq2$ yields examples of \sst s having \sst\
matrices which are not invertible over $\Z$ but which still contain all
information about the invariant measure on the hull.

\subsubsection*{Penrose tilings}

Penrose tilings belong to the best known two dimensional tilings.
In the form most suitable for us they consists of two unilateral
triangles that have angles which are multiples of $\frac{\pi}{5}$.
A part of such a tiling is shown in figure~\ref{bb1}.
These tilings were invented by R. Penrose before the discovery of quasicrystals
\cite{She}. Today a three dimensional generalization of a Penrose tiling
\cite{Gar,KrNe,LeSt1} serves as an idealized model for the spatial structure of
 quasicrystals
which yield tenfold symmetric Bragg reflexes.
Among the vast literature on these
tilings we refer to \cite{GrSh} and, in relation to the physics of
quasicrystals, to the diverse articles collected in \cite{OsSt}.\ms

Originally a Penrose tiling was defined as a "puzzle" by so-called matching
conditions. Later on various methods to obtain such a tiling from $\Z^5$ or
the root lattice $A_4$ of $SU(5)$ were found. For us it is important that some
Penrose tilings, e.g.\ the fivefold symmetric ones, are invariant
under the \sst\ displayed in figure~\ref{b2}.
The set of all Penrose tilings is the hull of such an invariant tiling.
In fact, as the \sst\ is $\Sy$-primitive such a tiling is homogeneous
so that any Penrose tiling defines the same hull.
The \sst\ is locally invertible  and the corresponding
map $\jo$ has been first given by Robinson though he did not view it as a map
between topological spaces \cite{GrSh}.
Connes used this map to derive the algebra $\AF$ and computed the ordered
\kog\ and its image under $\tr_*$ \cite{Con}. His results are justified by
confirming that the \sst\ satisfies \bii. In fact,
a Penrose tiling  has the
property that the boundaries of the $n$-fold \fl s of the triangles furnish
for $n\geq 2$ local mirror axes. This means that any \ti\ inside an
$n$-fold \fl\ which has an edge lying on that boundary
does also occur in the tiling reflected at that edge. To
illustrate this, figure~\ref{bb1ul} shows the $2$-fold \fl\ of the smaller
triangle
together with part of the tiles which are forced by it.
Further repetition of the \sst\ procedure leads to local centres of fivefold
symmetries at the corners of the $n+2$-fold \fl s, i.e.\ there are fivefold
symmetric
\mi s at these corners.
Hence the \sst\ determines its border ($N$ can be taken to be $4$).

The orientational symmetry is generated by a rotation around
$\frac{\pi}{5}$ together with a reflection on axes along the boundaries
(edges) of the triangles, hence $|\Sy|=20$.
However there are no tenfold symmetric \mi s. \ms

As may be seen from figure~\ref{b2} which also contains $\Sigma$ and
the embedding graph
of the \sst\ its \sst\ matrix is given by
\be
\sigma = \left(\begin{array}{cc}
 2 & 1 \\ 1& 1  \end{array}\right)
\ee
and has \pfw\ $\tau=t^2$ with $t=\frac{1+\sqrt{5}}{2}$
(the inverse of the golden mean) and normalized left-\pfv\
$\nu =(t-1,2-t)$.
This yields
\be
tr_*K_0(\AF) = \Integer + t \Integer
\ee
and $K_0(\AF)=\Integer^2$, as $\sigma$ is invertible over $\Z$.

\section{$\A_\tl$ versus $\AF$}

The $AF$-algebra $\AF$ constructed with the help of the \sst\ could be used
to determine an invariant measure on the hull.
However $\AF$ differs from $\A_\tl^\Sy$.
In fact due to the existence of singular tilings the image of
$\Gr_\Sigma$ under $\joh^{-1}\!\times\joh^{-1}$ is not closed in the
topology of $\Gr$.
Nevertheless $\AF$ may be
used to show that  the measure $\mu$ yields a trace on $\A_\tl$
which coincides with the one from the Shubin formula.
This is so far only clear for one dimensional tilings with trivial $\Sy$ and
Cartesian products
thereof: we actually determined the unique ergodic measure above of which has
been shown by Bellissard \cite{Be1} that it leads to the right trace.
\bs

Let us come back to the formulation of $\AF$ as groupoid-\CA. For
$\xi,\xi'\in\p{n}{i}$
the functions
$e_{\xi,\xi'} : \Gr_\Sigma\rightarrow \Complex$
\be
e_{\xi,\xi'} (\gamma,\gamma') =
\left\{
\begin{array}{l}
1\: \hbox{ if }\:\exists\gamma'':\,\gamma=\xi\circ\gamma''\:
\hbox{ and }\:
\gamma'=\xi'\circ\gamma'' \\
0\: \hbox{ else}
\end{array}      \right.
\ee
generate a topological $*$-algebra, the $C^*$-hull of which is $\AF$.
In the representations corresponding to (\ref{225}) the generators
$\pi_\alpha(e_{\xi,\xi'})$ act on  wavefunctions $\phi\in\ell^2
(\Gr_\Sigma^\alpha)$ by (again writing shorter
$\phi(\gamma) =\phi(\alpha,\gamma)$)
\be                              \label{332a}
(\pi_\alpha(e_{\xi,\xi'})\phi)(\gamma) =
\left\{
\begin{array}{ll}
\phi(\xi'\circ\gamma'') & \hbox{if }
\:\exists\gamma'':\,\gamma=\xi\circ\gamma'' \\
0 & \hbox{else} .
\end{array}      \right.
\ee
The restriction on $\Gr^T$ of the mapping $(T,T')\mapsto (\jo(T),\jo(T')):\,
\Gr^T\rightarrow\Gr_\Sigma^{\jo(T)}$ is bijective, if $T$ is regular.
Therefore we may represent $\AF$ also on $\ell^2(\Gr^T)$:
\begin{eqnarray}                                        \label{3311}
 (\pi_T(e_{\xi,\xi'})\psi)(T') & := &
(\pi_{\jo(T)}(e_{\xi,\xi'})\psi\circ\jo^{-1})(\jo(T')) \\
& = & \left\{
\begin{array}{ll}
\psi(\jo^{-1}(\xi'\circ\gamma'')) & \hbox{if }
\:\exists\gamma'':\,\jo(T')=\xi\circ\gamma'' \\
0 & \hbox{else}
\end{array}      \right. \\
\label{5apa0}
 & = & \left\{
\begin{array}{ll}
\psi(S-(x_{\xi'}\!-\!\pk{n}{T'})) & \hbox{if }\:
\jo_n(T')=\xi\\ 0 & \hbox{else},
\end{array}      \right.
\end{eqnarray}
where, as for (\ref{5pa1}), $x_{\xi'}\in\Mi{n}{T'}\2$ denotes the \pkt\ of the
\ti\ encoded by $\xi'$.
For singular tilings (\ref{5apa0}) shall be the definition of a representation
of $\AF$.
Condition  $\jo_n(T')=\xi$ implies that
$\pi_T(e_{\xi,\xi'})$ is only a translation by $(x_{\xi'}\!-\!\pk{n}{T'})$
in case both, initial and final point of the translation lie in a common
$n$-fold \fl. In particular a translation crossing a boundary
of an $n$-fold \fl\ may only be in $\pi_T(\AF^{(m)})$ for $m>n$.
This shows that the elements of $\pi_T(\A_\tl^\Sy)$, which is generated by
\be
\tilde{e}_{M,x,x'} = \sum_{g\in\Sy}e_{g\cdot M, g\cdot x,g\cdot x'},
\ee
 may not be approximated in norm by elements of $\pi_T(\AF)$, i.e.\ the
inclusion
\be    \label{5apa1}
\pi_T(\AF) \subset\pi_T(\A_\tl^\Sy)
\ee
which is guaranteed by the local invertiblity of $\rho$
is a proper one.
But one has:
\bt
For regular $T$
\be
\overline{\pi_T(\AF)}^s = \overline{\pi_T(\A_\tl^\Sy)}^s
\ee
where $\overline{\:\cdot\:}^s$ denotes strong closure in the algebra of bounded
operators on a Hilbert space.
\et
{\em Proof:} For $x,x'\in M\2$ let
$${\cal P}^{(n)}_{M,x,x'}:=\{(\xi,\xi')\in\bigcup_i\p{n}{i}\times\p{n}{i}|
\exists T'\in\jo_n^{-1}(\xi):(M,x,x')\subset
(\Mi{n}{T'},\pk{n}{T'},x_{\xi'})\}$$
 $x_{\xi'}\in\Mi{n}{T'}\2$ denoting the \pkt\ of the
\ti\ encoded by $\xi'$ and define
\be
\pi_T(\tilde{e}_{M,x,x'})^{(n)} :=
\sum_{(\xi,\xi')\in {\cal P}^{(n)}_{M,x,x'}} \pi_T(e_{\xi,\xi'}) ,
\ee
the $n$'th approximant of $\pi_T(\tilde{e}_{M,x,x'})$. Denote by
$\psi_{T'}$, $T'\sim T$ with $\psi_{T'}(T'')=\delta_{T'T''}$ the usual
orthonormal basis of
$\ell^2(\Gr^T)$
and by $r_n$ the inner radius of $(\Mi{n}{T},\pk{n}{T})$.
 Define subspaces ${\cal H}^{(n)}$ of $\ell^2(\Gr^T)$ to be generated by
$\{\psi_{T-x}|x\in \Mi{n}{T}\2,|x|<r_n\!-\!r_{0}\}$ where $r_0$ shall
be the radius of the smallest ball that covers $M$.
Then regularity of $T$ is equivalent to the statement that
$\forall T'\sim T\:\exists n:\psi_{T'}\in
{\cal H}^{(n)}$. Now on all $\psi\in{\cal H}^{(n)}$ the $n$'th approximant
acts exact:
$\pi_T(\tilde{e}_{M,x,x'})^{(n)}\psi=\pi_T(\tilde{e}_{M,x,x'})\psi$.
Thus
$\pi_T(\tilde{e}_{M,x,x'})^{(n)}$ strongly converges to
$\pi_T(\tilde{e}_{M,x,x'})$.
  Therefore contains $\overline{\pi_T(\AF)}^s$ the generators of
$\pi_T(\A_\tl^\Sy)$.\eb\bs

The regularity of $T$ is an essential condition for the theorem. In the strong
closure of a
representation of the $AF$-algebra on a symmetric tiling there are no
translations which over cross  a symmetry axis. In particular the Laplace
operator is not even in the strong closure. One might hope that the
influence of the symmetry axes on the spectrum of a generic
$\Sy$-invariant operator may be show up in a more thorough analysis of the
relation between $\A_\tl$ and $\AF$ and their $K$-groups.\bs

The above discussed representation may be used to show that the trace on
$\AF$ does satisfy the Shubin formula
if a representation on a regular tiling is considered.
Let $T$ be a regular tiling and consider its \ap. Writing shorter
$\Lambda_m=\Mi{m}{T}$ one has
\be
\Ch_{\Lambda_m} \pi_T(\A_\tl^\Sy) \Ch_{\Lambda_m} =
\pi_T(\A_{i_m}^{(m)})
\ee
where $i_m=r(\jo_m(T))$ and
 $\A_{i_m}^{(m)}$
is the  $i_m$'th direct summand of the $m$'th approximation of  $\AF$,
c.f.\ (\ref{m0}). For brevity we write
$h_n^{(n+m)}=h^{(n+m-1)}\circ\cdots\circ h^{n}:\A^{(n)}\rightarrow\A^{(n+m)}$
for the embedding of the $n$'th approximant into the $n\!+\!m$'th and
identify the approximants with their image under the embeddings $h_n$ in
$\AF$. In particular $\tr|_{\A^{(n)}}=\tr^{(n)}$.
\bt
For $a\in\AF$
\be            \label{a01}
\lim_{m\rightarrow\infty}\frac{1}{|\Lambda_m\2|} \hbox{Tr}(\Ch_{\Lambda_m}
\pi_T(a)) = \tr(a) ,
\ee
$\tr$ being the unique normalized trace on $\AF$.
\et
{\em Proof:}
Let $\chi_{i_m}^{(m)}\in\AF$ be the  projection on $\A_{i_m}^{(m)}$.
Since any two traces on finite dimensional simple \CA s are proportional
\be
\frac{1}{|\Lambda_m\2|} \hbox{Tr}(\Ch_{\Lambda_m} \pi_T(a)) =
\frac{\hbox{Tr}(\pi_T(\chi_{i_m}^{(m)} a))}
{\hbox{Tr}(\pi_T(\chi_{i_m}^{(m)}))} =
\frac{\tr^{(m)}(\chi_{i_m}^{(m)} a)}{\tr^{(m)}(\chi_{i_m}^{(m)})} ,
\ee
and we need to show that the r.h.s.\ converges for $n\rightarrow\infty$
to $\tr(a)$, once $a\in\AF$.
As both sides of (\ref{a01}) are continuous in norm we may restrict
to $a\in\AF^{(n)}$ for large enough $n$. Decompose
$a=\sum_i a_i$ with $a_i\in\A_{i}^{(n)}$. Then, c.f.\ (\ref{m5}),
\be
\tr^{(n+m)}(\chi_{i_{n+m}}^{(n+m)} h^{(n+m)}_n(a))=
\tau^{1-n-m} \nu_{i_{n+m}}\sum_j \sigma^m_{\:i_{n+m}j} \hbox{Tr}(a_j) ,
\ee
which asymptotically behaves like
$\tau^{1-n} \nu_{i_{n+m}} \sum_{j} S^{-1}_{\:i_{n+m}1}S_{1j} \hbox{Tr}(a_j)$,
$S$ here denoting the transformation that diagonalizes $\sigma$, i.e.\
$S\sigma S^{-1} = diag(\tau,\cdots)$, c.f.\ (\ref{pla1}).
Therefore
\be
\lim_{m\rightarrow\infty}\frac{\tr^{(m)}(\chi_{i_m}^{(m)} a)}
{\tr^{(m)}(\chi_{i_m}^{(m)})} =
\lim_{m\rightarrow\infty}\frac{\tr^{(n+m)}(\chi_{i_{n+m}}^{(n+m)}
h^{(n+m)}_n(a))}
{\tr^{(n+m)}(\chi_{i_{n+m}}^{(n+m)})} =
\frac{\sum_{j} S_{1j} \hbox{Tr}(a_j)}{\sum_{j} S_{1j}} = \tr(a) ,
\ee
from which the statement follows.\eb\bs

This theorem not only shows that the r.h.s.\ of (\ref{11}) coincides on
elements of $\pi_T(\AF)$ with the
trace on $\AF$ but by ($\ref{234}$) it even implies that this trace extends to
$\A_\tl$.
Thus to show that Shubin's formula holds  (\ref{140}) has to be verified.
As already mentioned, the proof of (\ref{140}) given in the appendix of
\cite{Be1} for cases where $\tl\2$ may be identified with an amenable group
does not rely on the group structure but only on the fact that there is an
increasing chain $\{\Lambda_n\}_{n\geq 1}$ approximating $\tl$ such that
$\lim_n\frac{|b(\Lambda_n)\2|}{|\Lambda_n\2|}\rightarrow 0$.
Hereby $b(\Lambda_n)$ denotes the \saum\ of $\Lambda_n$.
For the convenience
of the reader we present the proof here:

Let $\{\Lambda_n\}_{n\geq 1}$ be as for the above theorem, in particular
$\lim_n\frac{|b(\Lambda_n)\2|}{|\Lambda_n\2|}\rightarrow 0$. If $E$ lies in a
gap
$\Ch(H\leq E)\in C(H)$ and by the Stone-Weierstra\ss\ theorem it suffices
to proof that
\be                                \label{5apa2}
\lim_{n\rightarrow \infty}
\frac{1}{|\Lambda_n\2|} \mbox{Tr}
(\Ch_{\Lambda_n}H^k-(\Ch_{\Lambda_n}H)^k) = 0 ,
\ee
for any natural $k$.
First observe that
\begin{eqnarray}
\Ch_{\Lambda_n}(H^k-(\Ch_{\Lambda_n} H)^k) &=&
\Ch_{\Lambda_n} H(1-\Ch_{\Lambda_n})H^{k-1}+\Ch_{\Lambda_n}
H\Ch_{\Lambda_n}(H^{k-1}-(\Ch_{\Lambda_n} H)^{k-1}) \nonumber \\
&=&\sum_{j=1}^{k-1}(\Ch_{\Lambda_n} H)^j(1-\Ch_{\Lambda_n})
H^{k-j}. \nonumber
\end{eqnarray}
Since
$$\frac{1}{|\Lambda_n\2|} |\mbox{Tr}((\Ch_{\Lambda_n} H)^j(1-\Ch_{\Lambda_n})
H^{k-j})|\leq
\frac{1}{|\Lambda_n\2|} \mbox{Tr}(\Ch_{\Lambda_n})
\|H(\Ch_{\Lambda_n} H)^{j-1}(1-\Ch_{\Lambda_n})H^{k-j}\|\leq \|H\|^k$$
$H\mapsto \lim_{n\rightarrow \infty}
\frac{1}{|\Lambda_n\2|} \mbox{Tr}
(\Ch_{\Lambda_n}H^k-(\Ch_{\Lambda_n}H)^k)$ is  continuous and  we may restrict
to
$H\in\pi_T(C_c(\Gr))$.
Then $C:=\max\{|M\2||(M,x)\subset (T,y),
\forall x'\in M\2: H_{xx'}\neq 0\}$
is finite, and
$$ |\mbox{Tr}((\Ch_{\Lambda_n} H)^j(1-\Ch_{\Lambda_n})H^{k-j})|\leq
|b(\Lambda_n)\2|C\max_{x\in\tl\2}|\sum_{x'\in\tl\2}(\Ch_{\Lambda_n} H)^j_{xx'}
H^{k-j}_{x'x}| .$$
Therefore
$$ |\lim_{n\rightarrow \infty}
\frac{1}{|\Lambda_n\2|} \mbox{Tr}
(\Ch_{\Lambda_n}H^k-(\Ch_{\Lambda_n}H)^k)|\leq \lim_{n\rightarrow \infty}
\frac{|b(\Lambda_n)\2|}{|\Lambda_n\2|}C^2
\sum_{j=1}^{k-1}\max_{x,x'\in\tl\2}|H^j_{xx'}||H^{k-j}_{xx'}|$$
and (\ref{5apa2}) follows from $\lim_n\frac{|b(\Lambda_n)\2|}{|
\Lambda_n\2|}\rightarrow 0$.\eb
\subsection{Substitution sequences}    \label{sec5}

A one dimensional \sst\ tiling, examples of which were already given in
\ref{s421}, is also called a \sst\ sequence or automatic sequence.
It is well known how to
obtain the ergodic measure on the hull of such a tiling even in cases
where the \sst\ is neither locally invertible nor satisfies \bii\
\cite{Q,BBG}.
However, in case the hull may be
identified  with a path space, the induced homeomorphism yields a nice example
of a Vershik transform and one may easily show that $K_0(\A_\tl)=K_0(\AF)$
as scaled ordered groups.
In fact the equality as ordered groups
is already established once one has shown that
$\varphi$ induces a Vershik transform on the path space: it is given by
Theorem~8.3 in \cite{HPS}.
In fact \sst\ sequences of locally invertible \sst s satisfying \bii\ yield
concrete examples of minimal dynamical systems -- recall that by Lemma~\ref{l5}
$(\Om,\varphi)$ is minimal -- in which the Vershik transform is rather
simple and the $K_0$-groups them self, too. In particular for the \A-type
generalizations of the Fibonacci chain with $2k+1$ being non prime this is of
importance as there the elements of the $K_0$-group yield a finer gap
labelling.
Hereby again different letters are supposed to represent different congruence
classes of punctured intervals so that $\Sy$ is trivial. \bs

To compute $K_0(A_\tl)$ in case
 $\rho$ is locally invertible and satisfies \bii\ so that $\Om\cong\pf$
the minimal dynamical system $(\pf,\varphi)$ --
the induced homeomorphism $\joh\circ\varphi\circ\joh^{-1}$ on $\pf$  is simply
denoted by $\varphi$, too -- is periodically approximated. $\pf$ carries the
structure of the self-similarity
in form of a chain of partitions of itself which become finer and finer:
 $\{U_\xi\}_{\xi\in\pf^{(n)}}$ is a partition of $\pf$ into closed disjoint
subsets, and it is finer than
$\{U_\xi\}_{\xi\in\pf^{(n-1)}}$ in the sense that any element of the latter
is a disjoint union of elements of the former.
This allows us
to approximate $\pf$ by $\pf^{(n)}$ and the
shift operation by some map $\varphi_n$ operating cyclically on
$\pf^{(n)}$
and to obtain the ordered $K_0$-group of $\A_\tl$ as an inductive
 limit of the directed system of
the $K_0$-groups related to the periodic approximants. \bs

Denote by
 $\p{n}{i}$ the set of paths of length $n$ which end at $i$ and by
$\Sigma^{(1)}_{ji}$ the set of edges from vertex $i$ to $j$.
$\p{n}{i}$ may be identified with a subset of
$\p{n+1}{j}$ through concatenation, in case $\Sigma^{(1)}_{ji}\neq\emptyset$:
Any edge $\k\in\Sigma^{(1)}_{ji}$ defines by $\xi\mapsto\xi\circ\k$ an
injective map from $\p{n}{i}$ into
\be
\p{n+1}{j} = \bigcup_{i\in\Sigma^{(0)}_{}}
 \bigcup_{\k\in\Sigma^{(1)}_{ji}} \p{n}{i}\circ\k ,
\ee
where $\p{n}{i}\circ\k = \{\xi\circ\k|\xi\in\p{n}{i}\}$.
This structure of embeddings is preserved by $\varphi$ in the following sense.
Let $\gamma\in\pf$ be any path starting at $i$.
Recall that a path $\xi\in\p{n}{i}$ encodes the location of the letter
$a_{s(\xi)}$ in $\rho^n(a_{r(\xi)})$ and that $\varphi$ acts on tilings as
left shift. Thus the first $n$ edges of $\varphi(\xi\circ\gamma)$ encode the
location of the letter to the right of that $a_{s(\xi)}$ which was encoded by
$\xi$. Moreover, there is a unique path for which  $\varphi(\xi\circ\gamma)
\notin\p{n}{i}\circ\gamma$.
It is the path which  encodes the location of the last letter
 in $\rho^n(a_i)$; it
shall be denoted by $\gr{n}{i}$. Define
$\varphi_n : \pf^{(n)}\rightarrow\pf^{(n)}$  on
 $\pf^{(n)}\backslash\{\gr{n}{i}|i\in\Sigma^{(0)}\}$
by
\be
\varphi_n(\xi)\circ\gamma  :=  \varphi(\xi\circ\gamma)
\ee
and on $\{\gr{n}{i}|i\in\Sigma^{(0)}\}$ by
\be
\varphi_n(\gr{n}{i})  :=  \kl{n}{i} ,
\ee
$\kl{n}{i}$ being the path of  $\p{n}{i}$ which encodes the position of the
first letter in $\rho^n(a_i)$.
As a result $\varphi_n|_{\p{n}{i}}$ is cyclic of order $N_i^{(n)} =
|\p{n}{i}|$,
the number of paths in $\p{n}{i}$.

Stated differently $\varphi$ defines an order on $\p{n}{i}$, namely
$$\kl{n}{i} < \varphi_n(\kl{n}{i}) <\cdots <\varphi_n^{N_i^{(n)}-1}(\kl{n}{i})=
\gr{n}{i}$$ which corresponds to the order of the letters in $\rho^n(a_i)$.
The self-similarity of the tiling now reflects in the fact that these orders
are already determined by the orders on $\p{1}{i}=\bigcup_j\Sigma^{(1)}_{ij}$.
To see this consider the action of
$\varphi_{n+1}:\pf^{(n+1)}\rightarrow\pf^{(n+1)}$, i.e.\ on elements of the
form
$\xi\circ\k$ with $\xi\in\p{n}{s(\k)}$.
If $\xi\neq\gr{n}{s(\k)}$ then the letter encoded by $\xi\circ\k$ is not the
last letter of an $n$-fold \fl\ and hence neither the
last of an $n\!+\!1$-fold \fl. Thus for $\xi\neq\gr{n}{s(\k)}$
\be
\varphi_{n+1}(\xi\circ\k)\circ\gamma = \varphi(\xi\circ\k\circ\gamma)
=\varphi_n(\xi)\circ\k\circ\gamma .
\ee
Now, because of $\rho^{n+1}(a)=\rho^n(\rho(a))$ the last letter of
$\rho^n(\rho(a)_\nu)$ -- here $\rho(a)_\nu$ denotes the $\nu$'th letter of
$\rho(a)$
-- is mapped by $\varphi_{n+1}$ onto the first letter of $\rho(a)_{\nu'}$
where $\nu'=1$ if $\nu=N_{r(\k)}^{(n)}$ and $\nu'=\nu+1$ otherwise. In formula
\be
\varphi_{n+1}(\gr{n}{s(\k)}\circ\k) =
\kl{n}{s(\varphi_1(\k))}\circ\varphi_1(\k)
\ee
which shows by induction that $\varphi_n$ is determined by the order on the
$\p{1}{i}$.
Such a transformation is called {\em Vershik transform} as it has been
introduced by
Vershik \cite{Ver}. It is used in  \cite{HPS} for the analysis of
arbitrary minimal dynamical systems (over totally disconnected compact spaces).
A chain of partitions becoming finer and finer may then always be constructed
but it does in general not lead to a directed system of $K_0$-groups
which is stationary.

Let us assume for a moment that the \sst\ $\rho$ satisfies \bii\ in the
possibly stronger version
-- as for the examples given -- for which the first letter
$a_f$ as well as
the last letter $a_l$ of $\rho(a)$ are independent of $a$.
Then there is a unique minimal and a unique maximal path $\xi_{min}$,
$\xi_{max}$ in the sense of
\cite{HPS} which is also characterized by the property
$\varphi(\xi_{max})=\xi_{min}$ but $(\xi_{min},\xi_{max})\notin \Gr_\Sigma$.
In fact $\xi_{min}=\kl{1}{a_f}\circ\kl{1}{a_f}\circ\cdots$ and
$\xi_{max}=\gr{1}{a_l}\circ\gr{1}{a_l}\circ\cdots$. The singular tilings are
prescisely the elements of the $\Z$-orbit of $\jo^{-1}(\xi_{min})$,
 and
$U_{a_l a_f,1,2}$ contains $(\jo^{-1}(\xi_{max}),\jo^{-1}(\xi_{min}))$ as
accumulation point of points of $(\jo^{-1}\!\times\jo^{-1})\Gr_\Sigma$.
Hence $(\jo^{-1}\!\times\jo^{-1})\Gr_\Sigma$ is not a
closed subset of $\Gr$ in the topology of the latter.
$\Gr$ is in fact generated as an
equivalence relation by $(\jo^{-1}\!\times\jo^{-1})\Gr_\Sigma$
and the element $(\jo^{-1}(\xi_{max}),\jo^{-1}(\xi_{min}))$.
Theorem~8.3 of \cite{HPS}
implies $K_0(\A_\tl)\cong K_0(\AF)$ as
ordered groups. But let us add a purely $K$-theoretic computation which shows
that the above $K_0$-groups are isomorphic as groups their order isomorphy
then following from the results of the last section. It only requires
$\rho$ to satisfy \bii.\ms

Identifying the characteristic functions
$\Ch_{\xi}$ on $\xi\in \p{n}{i}$,
which are the generators of $C(\p{n}{i},\Z)$,
with the standard basis $\{e_i\}_i$ of $\Integer^{N_i^{(n)}}$
furnishes an isomorphism
\be       \label{301}
C(\p{n}{i},\Integer) \cong \Integer^{N_i^{(n)}}.
\ee
Let
$\Lambda_{N_i^{(n)}}$ be the ${N_i^{(n)}}\!-\!1$ dimensional
sublattice of $\Z^{N_i^{(n)}}$ generated by
$\alpha_i=e_i-e_{i+1}$. Then by the same identification
$E_n := \{f-f\circ \varphi_n^{-1}|f\in C(\pf^{(n)},\Integer)\}
\cong\Lambda_{N_i^{(n)}}$ and therefore
$C(\p{n}{i},\Integer)/E_n \cong \Integer^{N_i^{(n)}}/\Lambda_{N_i^{(n)}}
\cong \Integer$ or
\be
C(\pf^{(n)},\Integer) / E_n \cong \Integer^{|\Sigma^{(0)}|}.
\ee
As $\Ch_{\xi} \sim_{E_n} \Ch_{\xi'}$ whenever $r(\xi)=r(\xi')$,
 the embedding of groups
$\imath^{(n)} : C(\pf^{(n)},\Integer)/E_n  \rightarrow
C(\pf^{(n+1)},\Integer)/E_{n+1}$
\be
\imath^{(n)} [\Ch_\xi] :=
 [\Ch_{ \{\xi\circ\k | s(\k)=r(\xi)\} }] =
 \sum_{i\in\Sigma^{(0)}}\sum_{\k\in\Sigma^{(1)}_{i,r(\xi)}} [\Ch_{\xi\circ\k}]
\ee
is well defined. Hence $(C(\pf^{(n)},\Integer)/E_n,\imath^{(n)})$ is a
directed system of groups.
Using the isomorphism of (\ref{301}) one obtains the commuting diagram
\be                                       \label{3115}
\begin{array}{ccc}
C(\pf^{(n)},\Integer)/E_n & \stackrel{\imath^{(n)}}{\longrightarrow} &
C(\pf^{(n+1)},\Integer)/E_{n+1} \\
\downarrow & & \downarrow \\
\bigoplus_{i\in\Sigma^{(0)}}\Integer^{N_i^{(n)}}/\Lambda_{N_i^{(n)}} &
\stackrel{\sigma}{\longrightarrow} &
\bigoplus_{i\in\Sigma^{(0)}}\Integer^{N_i^{(n+1)}}/\Lambda_{N_i^{(n+1)}}
\end{array}
\ee
which shows that $(C(\pf^{(n)},\Integer)/E_n,\imath^{(n)})$
is as a directed system isomorphic to
$(\Integer^{|\Sigma^{(0)}|},{\sigma})$ the latter having algebraic limit
$K_0(\AF)$.
Recall that $K_0(\A_\tl)=C(\pf,\Integer)/E_\varphi$.
\bt
The direct algebraic limit of the directed system
$(C(\pf^{(n)},\Integer)/E_n,\imath^{(n)})$ is
\be
\lim_{\longrightarrow}
C(\pf^{(n)},\Integer)/E_n =
C(\pf,\Integer)/E_\varphi
\ee
where $\imath_n[\Ch_{\xi}] = [\Ch_{U_{\xi}}]$
for $\xi \in \pf^{(n)}$  and hence $K_0(\A_\tl)\cong K_0(\AF)$
as scaled ordered groups.
\et
{\em Proof:}
$\Ch_{\xi} \sim_{E_n} \Ch_{\xi'}$ implies that
$\exists k\in\Integer:\,\varphi^{k}(\xi) = \xi'$ and hence
$\varphi^{k}(U_{\xi}) = U_{\xi'}$. But
 $\exists k\in\Integer:\,\varphi^{k}(U_{\xi}) = U_{\xi'}$ implies
$\Ch_{U_{\xi}}\sim_{E_\varphi} \Ch_{U_{\xi'}}$.
Thus $\imath_n$ is well defined on equivalence classes.
Moreover, as
\be
\imath_{n+1}\circ\imath^{(n)} [\Ch_{\xi}]  =
\imath_{n+1} [\Ch_{\{\xi\circ \k |
 r(\xi) = s(\k)\}}]
 =  \sum_{\k\in\Sigma^{(1)},\,r(\xi) = s(\k)}[\Ch_{U_{\xi\circ \k}}] =
[\Ch_{U_{\xi}}]
\ee
diagram (\ref{m2}) commutes, namely $\imath_{n+1}\circ\imath^{(n)}=\imath_{n}$.
Finally $[\Ch_{U_{\xi}}]\mapsto
[(0,\cdots,0,\stackrel{|\xi|'te}{[\Ch_\xi]},0,\cdots)]$
is an isomorphism onto the standard realization of the inductive limit, c.f.\
(\ref{pla2}). This shows that $K_0(\A_\tl)\cong K_0(\AF)$
as groups.

To see that $K_0^+(\A_\tl) = K_0^+(\AF)$ recall
that
$\tr_*K_0(\A_\tl) = \tr_*K_0(\AF)$
by Theorem~\ref{thmb1}. In view of (\ref{pla01}) it has to be
shown that $K_0^+(\A_\tl) = \{z\in K_0(\A_\tl)|\tr_*(z)>0\}\cup\{0\}$.
The inclusion "$\subset$" follows from the faithfulness and positivity of
$\tr$.
To prove the opposite inclusion let $t=\tr_*(x)$ for some $x\in K_0(\A_\tl)$.
Then there is as well a $z\in K_0(\AF)$ with $\tr_*(z)=t$ and by the properties
of an $AF$-algebra  -- c.f.\ (\ref{pla02}) and the preceding remark --
a projection $p\in C(\pf)\otimes\K$ such that $\tr_*(z)=\tr(p)$.
By the identification $C(\Om)\cong C(\pf)$
of maximal commuting subalgebras of $\A_\tl$
and  $\AF$,
 which preserves the restrictions of the traces, the class of $p$ identifies
with a positive element of $K_0(\A_\tl)$ whose image under $\tr_*$ is $t$.

Finally $[1_{\A_\tl}]=[1_{\AF}]$ is directly seen from the
commuting diagram (\ref{3115}), since
 $[1_{\A_\tl}]=[\Ch_\Om]=[\Ch_{\pf}]$.\eb

\section*{Summary}
\addcontentsline{toc}{section}{\bf Summary}

An algebra which is well suited for the gap labelling of \So s on a non
periodic tiling has been defined purely from the geometrical data of the
tiling.
For its definition we discussed the hull of the tiling, its \nc\ space which is
a non Hausdorff set,
and the groupoid induced by the translations.
A topology on the groupoid was
found so that the algebra
$\A_\tl$ of the tiling $\tl$ could be defined as
the corresponding (reduced) groupoid \CA.
We restricted to a class of
tilings which lead to compact hulls.

It was shown that the gap labelling by means of the values of
the integrated density
of states is partly determined by an invariant measure on the hull
of the tiling. More specifically, for Cartesian products of one dimensional
systems the abstract gap labelling was solved
by expressing the \kg s of $\A_\tl$
as well as the image of $K_0(\A_\tl)$ under
$\tr_*$ through the ones of their one dimensional components. It turned out
that the invariant measure on the hull already fully determines
that image under $\tr_*$.

To obtain concrete results we specified to selfsimilar tilings which are
invariant under a locally invertible \sst\ satisfying an extra condition
\bii. For these a homoemorphism between the hull and a space of paths on a
graph could be constructed, which displays the topological structure in a very
clear way. Such a path space naturally defines a topological groupoid
$\Gr_\Sigma$ which
 identifies with a subset of the original groupoid $\Gr$ and induces
the same invariant measure. This measure was determined by the $K$-theory of
the $AF$-algebra
of the path space $\AF$, which is the groupoid \CA\ of $\Gr_\Sigma$.
Moreover the $AF$-algebra was used to show the validity of the Shubin
formula for \sst\ tilings.

Although $\Gr_\Sigma$ cannot be identified with a closed subset of $\Gr$
both algebras, $\AF$ and $\A_\tl$, have a lot in common.
We conjecture that $\tr_*K_0(\A_\tl)=\tr_*K_0(\AF)$ does not
only hold for Cartesian products of one dimensional \sst\ tilings but also for
general ones. One dimensional \sst\ tilings (sequences) are particularly
simple:
by a purely $K$-theoretic calculation it was shown that
the scaled ordered $K_0$-groups of $\A_\tl$ and $\AF$ coincide, their
difference being given by their $K_1$-groups.\bs

There are two dimensional substitution tilings with $8$-, and $12$-fold
orientational symmetry, e.g.\ the Ammann-Beenker and Socolar tilings,
which are of great interest, but which cannot be
treated yet in the above manner, as
their \sst s do not satisfy \bii.
An extension of the analysis of this paper to \sst s satisfying
weaker conditions is presently under investigation and will be the subject of a
future publication.

\newpage
\begin{appendix}
\section{Some $K$-theory and $AF$-algebras}

We briefly define the $K$-groups of a unital \CA\ referring the reader to
\cite{Bla} and \cite{Mur} for more general treatments. We apply it to
$AF$-algebras as an example which is of importance in the main text.
But first we recall the definition of the direct algebraic limit of a
directed system \cite{Eff,Lan,Mur}.\bs

A directed system is a family $(\A^{(n)},h^{(n)})$, $n\in\Z^{>0}$,
of objects and morphisms of a category
\be                       \label{M1}
\A^{(1)} \stackrel{h^{(1)}}{\longrightarrow} \A^{(2)}
\stackrel{h^{(2)}}{\longrightarrow} \cdots .
\ee
Its (direct or inductive) algebraic limit is categorically defined as
universal repelling object $(\A,h_n)$
for which the diagram
\be                                 \label{m2}
\begin{array}{rcl}
\A^{(n)} & \stackrel{h^{(n)}}{\longrightarrow} & \A^{(n+1)}  \\
h_n \searrow &  & \swarrow h_{n+1}  \\
 &\A&
\end{array}
\ee
commutes. It is often simply denoted by $\A=\lim_\rightarrow \A^{(n)}$.
For the category of Abelian groups the algebraic
limit exists and a
realization is given through the direct sum of all
 $\A^{(n)}$ modulo the equivalence relation which is generated by
\be \label{pla2}
(0,\cdots,0,\stackrel{n'th}{a},0,\cdots)\sim
(0,\cdots,0,\stackrel{n+1'th}{h^{(n)}(a)},0,\cdots).
\ee
For  \CA s it is the $C^*$-algebraic limit which is of interest. This is the
$C^*$-envelope of the above algebraic limit in which case we write instead
$\A = \overline{\lim_{\longrightarrow} \A^{(n)} }$. Any element of $\A$ may
be approximated by elements of the approximants, i.e.\ elements of the form
$h_n(a)$ with $a\in\A^{(n)}$ for some $n$.

\subsection{Preliminary $K$-theory of unital \CA s}

Let $\A$ be a unital \CA. $Proj(\A)$ denotes the set of equivalence classes of
projections of $\A$ under $p\sim q$ whenever $\exists u\in\A:p = uu^*,q =
u^*u$.
In order to introduce the structure of addition on $Proj(\A)$ the algebra
$\A$ has to be stabilized: Embed $M_k(\A)$
 (non-unitally) by  $*$-homomorphisms
\be
a\mapsto
\left( \begin{array}{cc}
a & 0\\
0 & 0
\end{array}\right)
\ee
into $M_{k+1}(\A)$.
The stabilization of $\A$ is the algebraic limit of the directed system so
obtained.
It is denoted by
$M_\infty(\A)$ and isomorphic to $\A\otimes\K$.
The completion of
$\K$ is isomorphic to the algebra of compact operators on a separable
Hilbert space.
Now the sum of
two projection classes $[p]$, $[q]$ of $V(\A):=Proj(M_\infty(\A))$ may be
defined by
\be
[p]+[q] =
\left[\left( \begin{array}{cc}
p & 0\\
0 & q
\end{array}\right)\right]
\ee
thereby $V(\A)$ becoming a monoid.
$K_0(\A)$, the $K_0$-group of $\A$, is obtained by Grothen\-dieck's
construction.
It is the quotient
$K_0(\A)=V(\A)\times V(\A)/\sim$ under
$([p],[q])\sim ([p'],[q'])$ whenever
$\exists [r]\in V(A):[p]+[q']+[r] = [q]+[p']+[r]$. \ms

In the cases of interest for us
in which $\A$ is unital and carries a (positive) faithful normalized
trace
$K_0^+(\A) = \{([p],[0])|\,[p]\in V(\A)\}$ defines a positive cone of
 $K_0(\A)$, i.e.\ a subset which is closed under addition and satisfies
$K_0^+(\A)-K_0^+(\A)=K_0(\A)$ as well $K_0^+(\A)\cap -K_0^+(\A)=\{0\}$.
In other words $K_0(\A)$ is an ordered group. To simplify the notation one
usually writes
$[p]$ in place of $([p],[0])$.
\ms

In order to distinguish different algebras having the same stabilization one
keeps track of the unit of $\A$, i.e.\ its $K_0$-class $[1]$ serves as
distinguished order unit. The set $\{[p]\in K_0^+(\A)|[0]\leq [p]\leq [1]\}$
defines a scale.
It coincides with the image of $Proj(\A)$ in $K_0^+(\A)$ if $\A$ has
cancellation, i.e.\ if $p\sim q,p\bot p',
q\bot q' \Rightarrow p'\sim q'$ holds for projections in $V(\A)$.
$K_0(\A)$ together with the above ordering and the distinguished order unit
$[1]$ is referred to as the scaled ordered $K_0$-group of $\A$.\bs

To define the  $K_1$-group of $\A$ one considers the groups of invertible
elements $GL_n(A)$ of $M_n(\A)$. These form under the embedding\be
a\rightarrow
\left( \begin{array}{cc}
a & 0\\
0 & 1
\end{array}\right)
\ee
a directed system of groups the algebraic limit of which is denoted by
$GL(\A)$. Let $GL(\A)_0$ be the connected component of $1$ (in the inductive
limit topology).
Then $K_1(\A) := GL(\A)/GL(\A)_0$ is as well an Abelian group the
multiplication being
\be
[u][v] = [uv] \sim
\left[\left( \begin{array}{cc}
u & 0\\
0 & v
\end{array}\right)\right] .
\ee
\bs

The computability of $K$-groups of \CA s is based on the fact that both, $K_0$
and $K_1$ are functors from the category of \CA s to the category of Abelian
groups which preserve direct sums and algebraic limits. Any  $*$-homomorphism
$h:\A\rightarrow \B$
induces a $*$-homomorphism $h\otimes 1:\A\otimes\K\rightarrow \B\otimes\K$
which
is usually also denoted by $h$ and a
group homomorphism
$h_*= K_i(h):K_i(\A)\rightarrow K_i(\B)$ through $h_*[x] = [h(x)]$. Similarly
a normalized trace $\tr$ on $\A$ induces a trace $\tr\otimes\mbox{Tr}:
\A\otimes\K\rightarrow \Complex$,
again usually denoted simply by $\tr$, and a state on $K_0(\A)$ by $\tr_*[x]=
\tr(x)$.
Moreover for a large class of \CA s the behavior of the functors under tensor
products is known. Namely if one restricts to nuclear \CA s, for which a unique
$C^*$-tensor product exists, the K\"unneth formula, which we cite in a
simplified way from \cite{Bla}, yields:
\bt
Let $\A,\B$ nuclear \CA s having \kg s which do not contain nilpotent elements.
Then
\begin{eqnarray} \label{ab1}
K_0(\A\otimes\B)& \cong& K_0(\A)\otimes K_0(\B)\:\oplus\:K_1(\A)\otimes
K_1(\B) \\                \label{ab2}
K_1(\A\otimes\B)& \cong& K_0(\A)\otimes K_1(\B)\:\oplus\:K_1(\A)\otimes
K_0(\B).
\end{eqnarray}
\et
Commutative \CA s and $AF$-algebras are nuclear as well as crossed products of
$\Z^d$ with nuclear \CA s.

\subsubsection*{$AF$-algebras}

$AF$-algebras belong to the simplest non-commutative algebras of infinite
dimension. They are the norm closure of the algebraic limit of a
directed system of finite dimensional
\CA s $\A^{(n)}$ with $*$-homomorphisms $h^{(n)}$.
Any finite dimensional \CA\ is semi-simple, i.e.\ it is a direct sum of simple
\CA s, and its direct summands (simple components) are isomorphic to
matrix algebras $M_m(\Complex)$. Hence the approximants
$\A^{(n)}$ of $\A$ are of the form
\begin{equation}             \label{m0}
\A^{(n)} =\bigoplus_{i=1}^{k_n} \A_i^{(n)}, \quad
\A_i^{(n)}\cong M_{N_i^{(n)}}(\Complex) .
\end{equation}
By the functoriality of $K_i$ one immediately obtains a directed system of
\kg s $(K_i(\A^{(n)}),h^{(n)}_*)$ whose algebraic limits furnish
the \kg s of $\A$. Because of $K_1(\Complex) = 0$, the system of $K_1$-groups
is trivial as well as its algebraic limit. On the other hand
 $K_0(\Complex) = \Integer$ implies that
$K_0(\A^{(n)}) = \Integer^{k_n}$, and
$h^{(n)}_*$ is a positive integer
$k_{n+1}\times k_n$ matrix, its $ji$-coefficient being given by the number of
times component
 $M_{N_i^{(n)}}(\Complex)$ is mapped by $h^{(n)}$ into
$M_{N_j^{(n+1)}}(\Complex)$.
The positive cone of
 $K_0(\A)$ is then given by
\begin{equation}            \label{m1}
 K_0^+(\A) = \bigcup_{n\geq 1}h_{n\,*}(K_0^+(\A^{(n)})) .
\end{equation}
The directed system of algebras may be encoded in a Bratteli diagram. A
Bratteli
diagram consists of a set of weighted vertices $V$ which are grouped into
floors
$V_n$, i.e.\ $V=\cup_{n\geq 1} V_n$, and of oriented edges between successive
floors. The $i$'th vertex of $V_n$ is identified with the generator of
$K_0(\A_i^{(n)})$ and has as weight the square root of the dimension of
 $\A_i^{(n)}$, that is $N_i^{(n)}$. It is joint to the $j$-th vertex of
$V_{n+1}$
by $(h^{(n)}_*)_{ji}$ oriented edges. These edges together with $V_n$ and
$V_{n+1}$ furnish the (bipartite) graph of the embedding of
$\A^{(n)}$ into $\A^{(n+1)}$ and the matrix $h^{(n)}_*$ is called the
embedding
matrix.

A Bratteli diagram, which hence also may be seen as a sequence of embedding
graphs, determines the directed system of algebras (up to isomorphism) and may
be considered as an invariant of it. Hence it also determines the $AF$-algebra,
however it may happen that different Bratteli diagrams which are not
comparable by a simple algorithm yield the same $AF$-algebra. For this reason
Bratteli
diagrams are unsuitable to classify $AF$-algebras. This is achieved by
$K$-theory. Elliot proved \cite{Ell} that the scaled ordered $K_0$-group is a
complete invariant for $AF$-algebras.\ms

The directed system of $K_0$-groups is already determined through
the unweighted Bratteli diagram alone. If the homomorphisms
 $h^{(n)}$ are all
unital embeddings, then the dimensions of the  $\A^{(n+1)}_j$ are already
determined by the ones of the
 $\A^{(n)}_i$ together with the matrix $h^{(n)}_*$. In this case all
weights are already determined by the weights of $V_1$. The weights of $V_1$
yield exactly the information of the order unit, namely $[1]=w$ where
$w_i=N_i^{(1)}$.\ms

We now specify to systems for which $h^{(n)}$ are unital embeddings and
$K_0(\A^{(n)})$ and
$h^{(n)}_*$ independent of $n$, i.e. the directed system is of the form
$(\Integer^r,\sigma)$ (stationary).
In that case, which is the one
occuring in the main text, the floors of the Bratteli diagram may all be
identified with $V_1$ and
the edges between them with the edges between $V_1$ and $V_2$
so that the whole diagram may be encoded
in a finite weighted graph $\Sigma$.
Namely $\Sigma$ has vertices $\Sigma^{(0)}= V_1$ and (oriented)
edges $\Sigma^{(1)}$ equal to the edges between $V_1$ and $V_2$ and the weights
are the weights of $V_1$. In particular, $\sigma$ is the \cm\ of $\Sigma$.
We denote the
$AF$-algebra which is defined by the Bratteli diagram by
 $\A_{\Sigma,w}$ where $w$ is the weight vector or, if
$w_i=1$ for all $i$, simply by $\AF$.
\ms

The faithful normalized traces on unital $AF$-algebras do exactly correspond to
the states on
the scaled ordered $K_0$-group. If in addition $\sigma$ is primitive the trace
on $\AF$ is up to normalization unique and correspondingly there is only one
state on that group \cite{GHJ},\cite{Eff}:
Let $\tau$ be the
\pfw\ and $\nu$ be the left-\pfv\ of $\sigma=h_*^{(n)}$ normalized to
$(\nu,w)=1$ (Euclidean
scalar product), then
\be                                      \label{m5}
\tr^{(n)}(a)=\tau^{1-n}\sum_{i}\nu_i\hbox{Tr}(a_i) ,
\ee
defines a system of faithful normalized traces on the directed system
$(\A_{\Sigma,w}^{(n)},h^{(n)})$ where $\hbox{Tr}$ is the usual trace of
matrices.
This system is compatible with the embedding structure, i.e.\
$\tr^{(n+1)}\circ h^{(n)} = \tr^{(n)}$ and may therefore be extended to the
algebraic limit (and its norm closure).
It induces a state $\tr_*$ on the scaled ordered $K_0$-group, namely for
$z\in K_0(\A^{(n)})\cong\Z^r$:
\be
\tr_*(h_{n\,*}(z)) =  \tau^{1-n}(\nu,z)
\ee
and the image of that state
is given by
\be                                         \label{m6}
\tr_*(K_0(\A_{\Sigma,w})) = \{\tau^{1-n}(\nu,z)|n>0,z\in\Integer^r\}
\ee
which is an additive subgroup of $\Real$.
This group is a self-similar subset of $\Real$ as it is invariant under
multiplication by $\tau$ and by $\tau^{-1}$.
As any element of $\AF$ is approximated by elements of the finite dimensional
approximants, every projection of $\AF\otimes\K$ is equivalent to a projection
of $C(\pf^{(n)})\otimes M_m(\Complex)$ for some $n,m$.
Hence, if $x\in \tr_*(K_0^+(\AF))$ there exists a projection $p\in
C(\pf)\otimes
\K$ such that $x=\tr(p)$, and
\be         \label{pla02}
\tr_*(K_0(\AF))=\mu(C(\pf,\Z))
\ee
$\mu$ being the invariant measure corresponding to $\tr$.
Let $S$ be
a base transformation diagonalizing $\sigma$, i.e.\
$S\sigma S^{-1} = diag(\tau,\cdots)$. Then $S_{1i}$ resp.\ $S^{-1}{}_{i1}$
are the components of the unnormalized left- resp.\ right-\pfv\ and may
therefore be chosen to be strictly positive. As $\tau$ exceeds all other

eigenvalues
of $\sigma$ in modulus one obtains for $z\in \Z^r$
\be \label{pla1}
\lim_{n\rightarrow\infty} \tau^{-n}(\sigma^n(z))_i = S^{-1}_{i1}
\sum_j S_{1j}z_j
\ee
and therefore: $\sum_j S_{1j}z_j > 0$ whenever
$\exists n:\sigma^n(z)\in(\Z^{>0})^r$. As a result
\be           \label{pla01}
K_0^+(\AF) = \{z\in K_0(\AF)|\tr_*(z)>0\}\cup\{0\} .
\ee
If $(\nu,z) = 0 \:\Rightarrow\:z=0$ then
$K_0(\A_{\Sigma,w})$ is totally ordered and may be identified with its
image under $\tr_*$.\ms

The algebraic limit of the $K_0$-group itself is easily obtained if $\sigma$ is
an automorphism of $\Z^r$. In this case $\sigma^{1-n}$ identifies
$K_0(\AF^{(n)})$ with $K_0(\AF^{(1)})$ and therefore
$\lim_\rightarrow K_0(\A^{(n)}) = \Integer^r$. If $\sigma$ is not over $\Z$
invertible but still $\frac{1}{\det \sigma}\sigma^r\in Aut(\Integer^r)$, then
$\sigma^{-rn}=(\frac{1}{\det \sigma})^n(\frac{1}{\det \sigma}\sigma^r)^{-n}$
which leads to $K_0(\A_{\Sigma,w}) = (\Integer[\frac{1}{\det \sigma}])^r$
as algebraic limit. In case $\sigma$ is neither degenerate nor
$\frac{1}{\det \sigma}\sigma^r$ an automorphism we cannot give such an explicit
form for the $K_0$-group. Finally, in case $\sigma$ is degenerate so that the
dimension of $F=\mbox{im} (h^r)$ strictly smaller than $r$, one may use the
 directed system $(F,h^r|_F)$. It is isomorphic to
$(\Integer^r,h)$, since
 \be
\begin{array}{rccc}
\Integer^r & \stackrel{h^r}{\longrightarrow} &
\Integer^r & \stackrel{h^r}{\longrightarrow}  \\
h^r \searrow & &id \nearrow \mbox{  } \searrow h^r&   \\
 &F&\stackrel{h^r|_F}{\longrightarrow}&F
\end{array}
\ee
commutes. Hence $K_0(\AF)$ coincides with the algebraic limit of $(F,h^r|_F)$
which may, if possible, be computed as described above.
\end{appendix}

\subsection*{Acknowledgements}

I want to thank  Werner Nahm for many ideas, fruitful conversations,
and encouraging support.
I also wish to thank
%Andreas H\"uffmann,
Andreas Recknagel
and Michael R\"osgen for stimulating discussions.
Andreas van Elst I would like to thank for discussions which
lead to the application of the K\"unneth formula and Le Tu Quoc Thang
for explanations concerning the topology of the hull.
I gratefully acknowledge conversations with the theory group in T\"ubingen.

\listoffigures
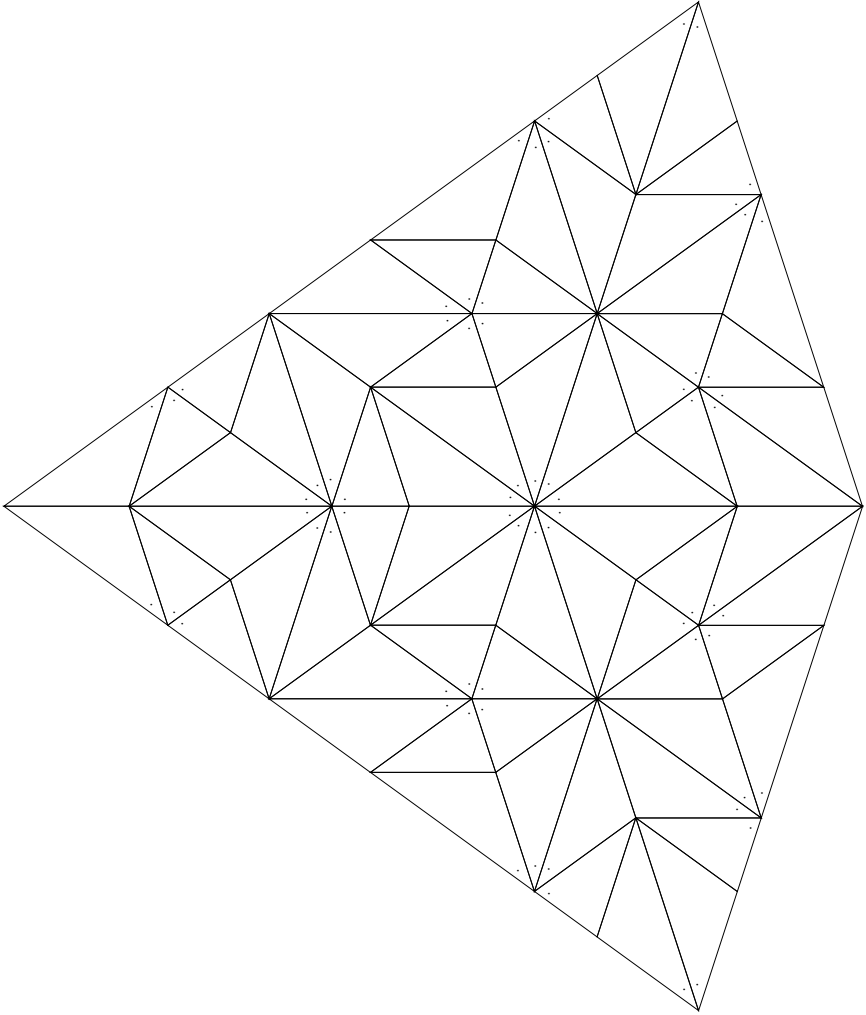
\begin{figure}[p]
\unitlength1cm
\begin{picture}(0,1)
\end{picture}
\caption[Part of a Penrose tiling]{\label{bb1}
Part of a Penrose tiling.}
\end{figure}

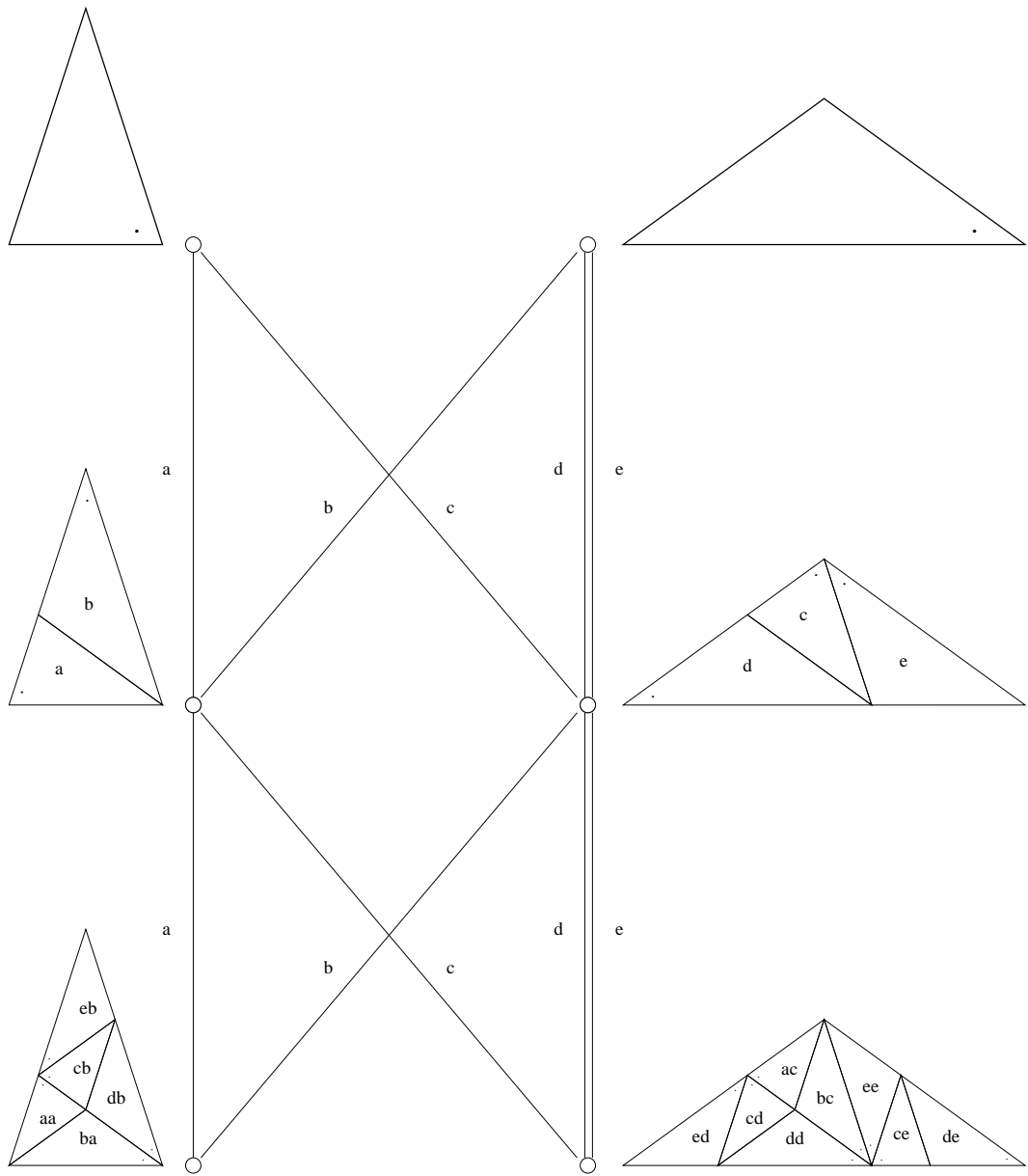
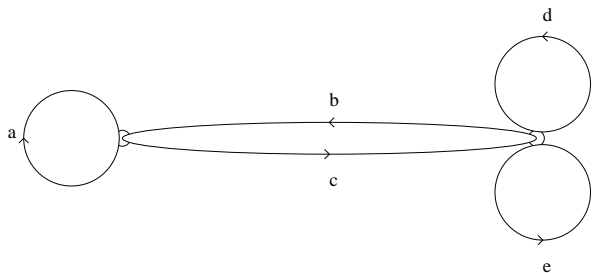
\begin{figure}[p]
\unitlength1cm
\begin{picture}(0,1)
\end{picture}
\caption[Deflation of the  Penrose tilings and Robinson map]
{\label{b2} Deflation of the Penrose tilings
together with an illustration for the Robinson map.
The graph $\Sigma$ is shown below.
Its (oriented) edges, which are
also the edges of the embedding graph, are denoted by a,b,c,d,e. The paths
of length $2$ encoding the positions of the tiles in the $2$-fold-\sst s are
indicated in the tiles.}
\end{figure}
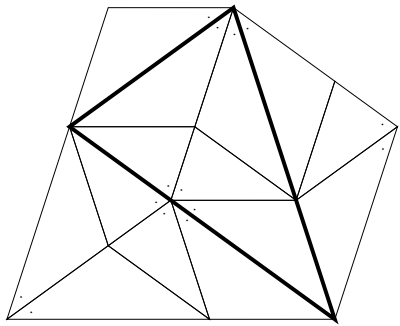
\begin{figure}[p]
\unitlength1cm
\begin{picture}(0,1)
\end{picture}
\caption[Local mirror axes of a Penrose pattern]{\label{bb1ul}
This is a pattern of a Penrose tiling which is determined by the $2$-fold
\sst\ of the smaller triangle. The boundaries of this $2$-fold \sst\ appear as
mirror axes.}
\end{figure}

\begin{thebibliography}{10}

\bibitem{Pen}
R.~Penrose.
\newblock The r\^ole of aesthetics in pure and applied mathematical research.
\newblock {\em Bull. Inst. Math. Appl.}, 10(7/8):266--71, 1974.

\bibitem{Gar}
M.~Gardner.
\newblock {\em Penrose Tiles to Trapdoor Ciphers}.
\newblock Freeman and Company (New York), 1989.
\newblock Here a three dimensional generalization of Penrose tilings which is
  due to R. Ammann is published.

\bibitem{KrNe}
P.~Kramer and R.~Neri.
\newblock On periodic and non-periodic space fillings of {$E^n$} obtained by
  projections.
\newblock {\em Acta Cryst. A}, 40:580--587, 1984.

\bibitem{LeSt1}
D.~Levine and P.J. Steinhardt.
\newblock Quasicrystals: A new class of ordered structures.
\newblock {\em Phys. Rev. Lett.}, 53(15):2477--2480, 1984.

\bibitem{She}
D.~Shechtman, I.~Blech, D.~Gratias, and J.V. Cahn.
\newblock Metallic phase with long range orientational order and no
  translational symmetry.
\newblock {\em Phys. Rev. Lett.}, 53:1951--1953, 1984.

\bibitem{BG}
A.~Bovier and J.M. Ghez.
\newblock Spectral properties of one dimensional {S}chr\"odinger operators with
  potentials generated by substitutions.
\newblock {\em Commun. Math. Phys.}, 158:45--66, 1993.

\bibitem{OsSt}
P.J. Steinhardt and S.~Ostlund.
\newblock {\em The Physics of Quasicrystals}.
\newblock World Scientific, 1987.

\bibitem{Con}
A.~Connes.
\newblock {\em {G\'eom\'etrie Non Commutative}}.
\newblock InterEditions (Paris), 1990.

\bibitem{GrSh}
B.~Gr\"unbaum and G.C. Shephard.
\newblock {\em Tilings and Patterns}.
\newblock Freeman and Company (New York), 1987.

\bibitem{Be1}
J.~Bellissard.
\newblock {$K$-Theory of $C^*$-Algebras in Solid State Physics}.
\newblock In T.C. Dorlas, N.M. Hugenholtz, and M.~Winnik, editors, {\em
  Statistical Mechanics and Field Theory: Mathematical Aspects}. Lect. Notes in
  Phys. 257:99-156, Springer Verlag, 1986.

\bibitem{Be2}
J.~Bellissard.
\newblock Gap labelling theorems for {S}chr\"odinger's operators.
\newblock In M.~Waldschmidt, P.~Moussa, J.M. Luck, and C.~Itzykson, editors,
  {\em From Number Theory to Physics}. 538-630, Springer Verlag, 1992.

\bibitem{Ver}
A.M. Ver\v{s}ik.
\newblock Uniform algebraic approximation of shift and multiplication
  operators.
\newblock {\em Soviet Math. Dokl.}, 24(1):97--100, 1981.

\bibitem{HPS}
R.H. Herman, I.F. Putnam, and C.F. Skau.
\newblock Ordered {B}ratteli diagrams, dimension groups and topological
  dynamics.
\newblock {\em Int. J. Math.}, 3:827--864, 1992.

\bibitem{BBG}
J.~Bellissard, A.~Bovier, and J.M. Ghez.
\newblock Gap labelling theorems for one dimensional discrete {S}chr\"odinger
  operators.
\newblock {\em Rev. Math. Phys.}, 4:1--38, 1992.

\bibitem{ReSi}
M.~Reed and B.~Simon.
\newblock {\em Methods of Modern Mathematical Physics}, volume~I.
\newblock Academic Press, 1972.

\bibitem{Be3}
J.~Bellissard.
\newblock Spectral properties of {Schr\"odinger's} operator with a {Thue-Morse}
  potential.
\newblock In J.M. Luck, P.~Moussa, and M.~Waldschmidt, editors, {\em Number
  Theory and Physics}. Springer Proceedings in Phys. 47:140-150, Springer
  Verlag, 1990.

\bibitem{BBG2}
J.~Bellissard, A.~Bovier, and J.M. Ghez.
\newblock Spectral properties of a tight binding {H}amiltonian with period
  doubling potential.
\newblock {\em Commun. Math. Phys.}, 135:379--399, 1991.

\bibitem{Luc}
J.M. Luck.
\newblock Cantor spectra and scaling of gap widths in deterministic aperiodic
  systems.
\newblock {\em Phys. Rev. B}, 39(5834-5849):596--616, 1989.

\bibitem{BKS}
M.~Baake, R.~Klitzing, and M.~Schlottmann.
\newblock Perfect matching rules for undecorated triangular tilings with 10-,
  12-, and 8-fold symmetry.
\newblock {\em Int. J. Mod. Phys. B}, 7:1455--1474, 1993.

\bibitem{Soc}
J.E.S. Socolar.
\newblock Simple octogonal and dodecagonal quasicrystals.
\newblock {\em Phys. Rev. B}, 39(15):10519--10551, 1989.

\bibitem{Ren}
J.~Renault.
\newblock {\em A Groupoid approach to $C^*$-Algebras}.
\newblock Lect. Notes in Math. 793. Springer-Verlag, 1980.

\bibitem{PiVo}
M.V. Pimsner and D.~Voiculescu.
\newblock Exact sequences for {$K$}-groups and {$Ext$}-groups of certain
  cross-product {$C^*$}-algebras.
\newblock {\em J. Operator Theory}, 4:93--118, 1980.

\bibitem{Put}
I.F. Putnam.
\newblock On the topological stable rank of certain transformation group
  {$C^*$}-algebras.
\newblock {\em Ergod. Th. {\&} Dynam. Sys.}, 10:197--207, 1990.

\bibitem{Tom}
J.~Tomiyama.
\newblock {\em Invitation to $C^*$-Algebras and Topological Dynamics}.
\newblock World Scientific, 1987.

\bibitem{Els1}
A.~van Elst.
\newblock {Anwendung nichtkommutativer Geometrie in der Spektraltheorie
  zweidimensionaler diskreter Schr\"odinger-Operatoren}.
\newblock Bonn Diplomarbeit BONN-IR-92-48, 1992.

\bibitem{Kel}
J.~Kellendonk.
\newblock On the algebraic characterization of aperiodic tilings related to
  {$ADE$}-root systems.
\newblock Bonn preprint BONN-HE-92-26, hep-th 9210078, 1992.

\bibitem{Els2}
A.~van Elst.
\newblock Gap-labelling theorems for {S}chr\"odinger operators on the square
  and cubic lattice.
\newblock Bonn preprint BONN-HE-93-26, to appear in Rev. Math. Phys.

\bibitem{Pi}
M.V. Pimsner.
\newblock Ranges of traces on {$K_0$} of reduced crossed products by free
  groups.
\newblock In {\em Operator Algebras and their Connections with Topology and
  Ergodic Theory}. Lect. Notes in Math. 1132:374-408, Springer Verlag, 1983.

\bibitem{Q}
M.~Queff\'elec.
\newblock {\em Substitution Dynamical Systems}.
\newblock Lect. Notes in Math. 1294. Springer-Verlag, 1987.

\bibitem{BGJ}
M.~Baake, U.~Grimm, and D.~Joseph.
\newblock Trace maps, invariants, and some of their applications.
\newblock {\em Int. J. Mod. Phys. B}, 7:1527--1550, 1993.

\bibitem{Bla}
B.~Blackadar.
\newblock {\em K-Theory for Operator Algebras}.
\newblock MSRI Publication 5. Springer Verlag, 1986.

\bibitem{Mur}
G.J. Murphy.
\newblock {\em $C^*$-Algebras And Operator Theory}.
\newblock Academic Press, 1990.

\bibitem{Eff}
E.G. Effros.
\newblock {\em Dimensions and $C^*$-Algebras}.
\newblock Conference Board Math. Sci. 46. Amer. Math. Soc. (Providence R.I.),
  1981.

\bibitem{Lan}
S.~Lang.
\newblock {\em Algebra}.
\newblock Addison-Wesley, 1984.

\bibitem{Ell}
G.~Elliot.
\newblock On the classification of inductive limits of sequences of semisimple
  finite dimensional algebras.
\newblock {\em J. Algebra}, 38:29--44, 1976.

\bibitem{GHJ}
F.M. Goodman, P.~de~la Harpe, and V.F.R. Jones.
\newblock {\em Coxeter Graphs and Towers of Algebras}.
\newblock MSRI Publication 14. Springer-Verlag, 1989.

\end{thebibliography}
\end{document}